\documentclass[12pt]{article}
\usepackage{amsmath,amsfonts,amssymb,amsxtra}
\usepackage{graphicx}
\input epsf

\makeatletter\@addtoreset{equation}{section}\makeatother

\renewcommand{\title}[1]{\vbox{\center\LARGE{#1}}\vspace{5mm}}
\renewcommand{\author}[1]{\vbox{\center#1}\vspace{5mm}}
\newcommand{\address}[1]{\vbox{\center\em#1}}
\newcommand{\email}[1]{\vbox{\center\tt#1}\vspace{5mm}}

\numberwithin{equation}{section}

\usepackage{hyperref}

\usepackage[numbers,sort&compress]{natbib}
\usepackage{hypernat}

\newcommand{\myref}[1]{}

\newcommand {\la} {\left \langle}
\newcommand {\ra} {\right \rangle}
\newcommand {\lb} {\left (}
\newcommand {\rb} {\right )}

\newcommand {\CalO} {\mathcal O}

\newcommand {\CalN} {\mathcal N}

\newcommand {\BR}   {\mathbb R}

\newcommand {\BC}   {\mathbb C}

\newcommand {\CP}   {\mathbb C \mathbb P}

\newcommand {\ve}  {\varepsilon}

\newcommand {\p} {\partial}

\DeclareMathOperator{\tr} {tr}

\DeclareMathOperator{\Pexp} {Pexp}

\newcommand{\SU}{SU}

\newcommand{\const}{\mathrm{const}}

\begin{document}

\unitlength = .8mm

\bibliographystyle{utphys}

\begin{titlepage}
\begin{center}
\hfill \\
\hfill \\

%\preprint{ITEP-TH-XX/09}

\title{Correlators of local operators and 1/8 BPS Wilson loops on $S^2$ from 2d YM and matrix models}

\renewcommand{\thefootnote}{\fnsymbol{footnote}}
\author{Simone Giombi$^{a}$ and
Vasily Pestun$^{b,\,}$\footnotemark}
\footnotetext{On leave of absence from ITEP, 117218, Moscow, Russia}

\address{Center for the Fundamental Laws of Nature
\\Jefferson Physical Laboratory, Harvard University,\\
Cambridge, MA 02138 USA\\
}

\email{$^a$giombi@physics.harvard.edu,
$^b$pestun@physics.harvard.edu}

\end{center}

\abstract{We propose that, in ${\cal N}=4$ Super Yang-Mills theory, correlation functions of certain 1/8 BPS Wilson loops and local operators inserted on a $S^2$ in space-time may be computed in terms of analogous observables in the ``zero-instanton" sector of 2d Yang-Mills theory. The Wilson loops are mapped to the standard Wilson loops of the 2d theory, as recently conjectured, while the local operators are mapped to powers of the 2d field strength. We give several perturbative checks of the correspondence, and derive from 2d Yang-Mills a two-matrix model for the correlator of a local operator and a Wilson loop of arbitrary shape. We show that the strong coupling planar limit of the two-matrix model precisely agrees with a string theory calculation in $AdS_5 \times S^5$.}

\vfill

\end{titlepage}

\eject

\tableofcontents

%%% Local Variables: 
%%% mode: latex
%%% TeX-master: "main"
%%% End: 

\section{Introduction}

Supersymmetric Wilson loops in the four-dimensional $\CalN=4$ super Yang-Mills theory have been extensively studied over the past years. One specific motivation is that, in certain cases, they may provide examples of physical observables which are non-trivial and yet exactly calculable. In particular, one may obtain this way interesting quantitative tests of the duality to type IIB string theory in the $AdS_{5} \times S^{5}$ background \cite{Maldacena:1997re,Witten:1998qj,Gubser:1998bc}. 

In \cite{Erickson:2000af,Drukker:2000rr} an exact result has been conjectured for the circular maximally supersymmetric 1/2 BPS Wilson loop operator: its expectation value can be computed using a Gaussian Hermitian matrix model. This conjecture has passed many subsequent tests, in particular it agrees with all the available calculations in the dual string theory\footnote{See \cite{Semenoff:2002kk} for a review of earlier work, and \cite{Drukker:2005kx,Gomis:2006sb,Gomis:2006im,Yamaguchi:2006tq,
Yamaguchi:2006te,Lunin:2006xr,D'Hoker:2007fq,Okuda:2007kh,Okuda:2008px} for a partial sample of more recent relevant work.}. In \cite{Pestun:2007rz} the Gaussian matrix model has been derived using localization of the four-dimensional path integral to supersymmetric configurations.

In \cite{Drukker:2007qr,Drukker:2007yx,Drukker:2007dw} a large class of interesting, generically $1/16$ BPS, Wilson loops has been found. Those loops live on a three-sphere $S^3$ in Euclidean space-time $\BR^4$. 
Imposing some restrictions on these Wilson loops on $S^3$ one gets various $1/8$, $1/4$ and $1/2$ BPS Wilson loops. In particular, restricting to the equator two-sphere $S^2 \subset S^3$, one gets generically $1/8$ BPS Wilson loops. In \cite{Drukker:2007qr,Drukker:2007yx,Drukker:2007dw} it has been conjectured that 
the expectation values of $1/8$ BPS Wilson loops on $S^2$ are exactly captured by a purely perturbative calculation in the two-dimensional bosonic Yang-Mills theory on $S^2$. In two dimensions, the preferred gauge choice is the light-cone gauge, since then there are no interactions. The conjecture of \cite{Drukker:2007qr,Drukker:2007yx,Drukker:2007dw} then implies that the 4d expectation values should be equal to the sum of the ladder diagrams of 2d Yang-Mills in light-cone gauge\footnote{More precisely, this is an Euclidean version of the light-cone gauge, defined by $A_{\bar z} = 0$, where $A$ is the 2d gauge field and $z,\bar z$ are complex coordinates on $S^2$.}. This prescription is not equivalent to the exact 2d bosonic Yang-Mills 
theory~\cite{Migdal:1975zf,Blau:1991mp,Blau:1993hj,Witten:1991we}, but instead to 
a ``truncation by hands'' of all non-zero instantons on $S^2$~\cite{Bassetto:1998sr,Bassetto:2001mf,Staudacher:1997kn}.
In \cite{Young:2008ed,Bassetto:2008yf} the conjecture has been supported at $\lambda^2$ order for a
single Wilson loop on $S^2$ (in the case of the ``two-longitudes" loop \cite{Young:2008ed,Bassetto:2008yf} as well as ``wavy-latitude" loops \cite{Young:2008ed}), where $\lambda = g_{YM}^2 N$ is the 't Hooft coupling constant for Yang-Mills theory with $\SU(N)$ gauge group. In particular, in \cite{Young:2008ed,Bassetto:2008yf} it was found that even when there are non-trivial $\CalN=4$ SYM interacting Feynman diagrams, the final result agrees with the ladder diagram computation in 2d YM in light-cone gauge. 
However, for a connected correlator of two latitudes on $S^2$, \cite{Young:2008ed} found a discrepancy at order $\lambda^3$ between the Feynman diagrams in $\CalN=4$ SYM and the ladder diagrams of 2d YM in light-cone gauge. Subsequently, in \cite{Giombi:2009ms} and \cite{Bassetto:2009rt} several new tests in support of the original conjecture have appeared. In particular it was shown in \cite{Giombi:2009ms} that invariance under are preserving diffeomorphisms holds at strong coupling, as implied by the conjecture, and a Gaussian two-matrix model for the connected correlator of two Wilson loops on $S^2$ was derived from the exact solution of 2d YM \cite{Giombi:2009ms}\cite{Bassetto:2009rt}. Its strong coupling limit agrees with the fact that there are no connected supersymmetric string worldsheets joining the two loops  \cite{Giombi:2009ms}, and the first subleading corrections to the saddle point at strong coupling have been shown to agree in \cite{Bassetto:2009rt} with the exchange of light supergravity modes between the two worldsheets. Further, in \cite{Bassetto:2009rt} a Feynman diagram calculation in ${\cal N}=4$ SYM at order $\lambda^3$, in the limit of one shrinking loop, was shown to be consistent with the two-matrix model. 

In \cite{Pestun:2009nn} the localization framework was used again to understand the relation
between the four-dimensional $\CalN=4$ SYM and the two-dimensional theory on $S^2$,
where the interesting Wilson loops live. Using localization, one gets naturally
a Lagrangian formulation of the conjectured 2d theory, which turns out to be 
closely related to 2d Hitchin/Higgs-Yang-Mills theory \cite{Moore:1997dj,Gerasimov:2006zt,Gerasimov:2007ap},
and also a natural explanation of the prescription to truncate the 2d instantons in the 2d YM conjecture. 
The localization computation in \cite{Pestun:2009nn} for 1/8 BPS loops was not completed at the same 
level of rigour as in \cite{Pestun:2007rz} for 1/2 BPS loops. One still needs to evaluate the
one-loop determinant for the field fluctuations in the directions normal to the localization locus. 
However, there are many reasons to believe that such determinant is trivial in the $\CalN=4$ theory,
and then the localization  \cite{Pestun:2009nn} would support the original conjecture \cite{Drukker:2007yx,Drukker:2007qr,Drukker:2007dw}. Moreover, the localization framework allows to establish a complete correspondence between 
all observables in the $\CalN=4$ SYM which share a number of certain superconformal symmetries 
and observables of the 2d theory. In particular, one immediate consequence is establishing 
the 2d description of certain local operators on $S^2$ which share 
some common superconformal symmetries with the relevant 1/8 BPS Wilson loops. These operators are chiral primaries equipped with an explicit space-time dependence, of the form $\tr(\frac{x^i}{r}\Phi_i+i\Phi_B)^J$, where $x^i$ are coordinates on $\mathbb{R}^3$ on which the two-sphere $x^ix^i=r^2$ is embedded, $\Phi_i$ are the three scalars which couple to the 1/8 BPS Wilson loops, and $\Phi_B$ is any of the remaining three scalars. Note that the definition involves an identification of a $SO(3)$ subgroup of the $R$-symmetry group with the $SO(3)$ rotating the $x^i$'s, as it is natural in the construction of the 1/8 BPS Wilson loops of \cite{Drukker:2007dw,Drukker:2007yx,Drukker:2007qr}. Actually, these local operators on $S^2$ are a special case of a more general class of protected operators on $\mathbb{R}^4$ which was recently studied in \cite{Drukker:2009sf}, where the preserved supersymmetries and the non-renormalization properties of their correlation functions were investigated. When an arbitrary number of such local operators is inserted on $S^2$, the system preserves 4 superconformal supercharges. On the other hand, the combined system of Wilson loops and local operators on $S^2$ preserves 2 common supercharges, which is sufficient for the localization of \cite{Pestun:2009nn} to be applicable. In particular, it follows that these local operators are mapped in the 2d theory to insertions of powers of the YM field-strength (or more precisely its Hodge dual). 

In this note we make several detailed weak and strong coupling tests of this correspondence involving
the chiral primaries on $S^2$, and our results support the 2d YM conjecture. In particular we study, to leading order in perturbation theory, the correlator of a local operator and a Wilson loop and obtain agreement with the corresponding computation in 2d YM. Further, from summing up the ladder diagrams of 2d YM in light-cone gauge we derive a two-matrix model for the the exact correlator of a local operator and a Wilson loop. This can also be written as a complex matrix model, and then one can see that our results imply as a special case the original conjecture of \cite{Semenoff:2001xp} for the exact correlator of a 1/2 BPS circular loop and a chiral primary\footnote{This was extended to the 1/4 BPS circular loop in \cite{Semenoff:2006am}. See also \cite{Semenoff:2002kk,Zarembo:2002ph,Pestun:2002mr,Okuyama:2006jc,Giombi:2006de,Gomis:2008qa} for additional related work}. We solve the two-matrix model in the planar limit, and show that at strong coupling it exactly matches the corresponding string theory calculation in $AdS_5\times S^5$, using the explicitly known string solutions for the 1/4 BPS latitude \cite{Drukker:2006ga}\cite{Drukker:2007qr} and the 1/4 BPS two-longitudes \cite{Drukker:2007qr} loops.  

%Hence, until now, all available results essentially support the original 2d YM conjecture,
%except the result in \cite{Young:2008ed} on the connected correlator of two Wilson loops at $\lambda^3$ %order. In \cite{Bassetto:2009rt} an attempt has been made to resolve the discrepancy by proposing 
%that for the purpose of computing the connected correlator of two Wilson loops the 
%light-cone gauge prescription $A_{\bar z} = 0$ is NOT equivalent to the Hermitian two-matrix
%model~\cite{Giombi:2009ms,Bassetto:2009rt} which captures the zero-instanton sector of the 2d YM.
%In appendix to this note we argue that this explanation does not work. Actually, we show 
%that the ladder Feynman diagrams in the light-cone gauge $A_{\bar z} = 0$ 
%for the correlator of two latitude Wilson loops on $S^2$ are in 
%precise agreement with the Feynman diagrams of the Hermitian two-matrix model.

Hence, until now, all available results essentially support the original 2d YM conjecture,
except the result in \cite{Young:2008ed} on the connected correlator of two Wilson loops at $\lambda^3$ order. In an attempt to explain such discrepancy, in \cite{Bassetto:2009rt} doubts were raised as to whether the light-cone gauge prescription $A_{\bar z} = 0$ is equivalent to the Hermitian two-matrix
model~\cite{Giombi:2009ms,Bassetto:2009rt} which captures the zero-instanton sector of 2d YM on $S^2$.
In appendix we show that the ladder Feynman diagrams in the light-cone gauge $A_{\bar z} = 0$ 
for the correlator of two latitude Wilson loops on $S^2$ are actually in 
precise agreement, to all orders, with the Feynman diagrams of the Hermitian two-matrix model, so the discrepancy in \cite{Young:2008ed} still remains unsolved.  

The paper is organized as follows. In Section \ref{prelim-sec} we set up our notations and conventions, we explain the 2d description of the local operators implied by localization, and we study the supersymmetries preserved by local and Wilson operators on $S^2$. In Section \ref{pert-sec} we give our perturbative checks of the proposed 2d-4d correspondence. In Section \ref{2mm-sec} we derive the two-matrix model from 2d YM in light-cone gauge and solve it in the planar limit. In Section \ref{string-sec} we present the string theory calculation of the correlator between a local operator and a Wilson loop. Finally, in the Appendix we collect some notes about 2d YM in light-cone gauge and the equivalence with the Gaussian matrix models derived from the zero-instanton sector.

\section{Preliminaries}
\label{prelim-sec}
\subsection{Notations and conventions}
The $\CalN=4$ SYM action on $\BR^4$ with the standard flat metric is
\begin{equation}
  S_{SYM} = -\frac 1 {g_{YM}^2} \int d^4 x \lb \frac 1 2 \tr  F_{\mu \nu} F_{\mu \nu} + \tr D_{\mu} \Phi_A D_{\mu} \Phi_A + \dots \rb\,,
\end{equation}
where $\mu=1,\ldots,4$ are space-time indices and $A=1,\ldots, 6$ are $SO(6)_R$ indices. 
Here we use conventions such that the covariant derivative is $D = d + A$, the curvature is
$F_{\mu\nu}=[D_{\mu}, D_{\nu}]$, and all fields take value in the Lie algebra of the gauge group, $A_{\mu} = A^a_{\mu} T_a, \Phi=\Phi^a T_a$, e.g. in the anti-Hermitian matrices for the $U(N)$ gauge group. The anti-Hermitian generators satisfy $\tr T_{a} T_{b} = -\frac 1 2 \delta_{a b}$. Hence the action may be also written as 
\begin{equation}
  \label{eq:S4d}
  S_{SYM} = \frac 1 {2g_{YM}^2} \int d^4 x \lb \frac {1} {2} F^a_{\mu \nu} F^a_{\mu \nu} + D_{\mu} \Phi^a_A D_{\mu} \Phi^a_A + \dots \rb.
\end{equation}

The 1/8 BPS Wilson loops of \cite{Drukker:2007qr,Drukker:2007yx,Drukker:2007dw} are located on a sphere $S^2$ of radius $r$ defined as $x_4 = 0, \sum_{i=1}^{3} x_i^2 = r^2$, and they couple to three of the six scalars, $\Phi_i,\,i=1,2,3$\footnote{Here we use the conventions in \cite{Drukker:2007qr,Drukker:2007yx,Drukker:2007dw}. These differ from the conventions used in \cite{Pestun:2009nn} by a relative sign in the scalar couplings.}
\newcommand{\dimR}{d_R}
\newcommand{\dimRpr}{d_{R'}}
\begin{equation}
  \label{eq:Wilson-loop}
  W_{R} (C) = \frac{1}{\dimR}\tr_{R} \Pexp \oint_C (A_j  + i \ve_{ijk} \Phi_i
  \frac {x^k} {r} ) dx^j\,,
\end{equation}
where $\dimR$ denotes the dimension of the representation $R$. According to the conjecture of \cite{Drukker:2007qr,Drukker:2007yx,Drukker:2007dw}, this supersymmetric operator  
is mapped to the standard Wilson loop on the same contour $C$ and representation $R$ in the two-dimensional Yang-Mills theory with action 
\begin{equation}
  \label{eq:S_{2d}}
  S_{SYM} = \frac 1 {2g_{2d}^2} \int d^2 \sigma \sqrt{g}  \lb  \frac {1} {2} \tilde F^a_{\mu \nu} \tilde F_a^{\mu \nu} \rb\,,
\end{equation}
where 
\begin{equation}
  \label{eq:4d-2d-relation}
  g_{2d}^2 = - \frac {g_{4d}^2} {2 \pi r^2}.
\end{equation}

We will use the notation $\tilde A, \tilde F$ for the fields of the two-dimensional theory. In particular \cite{Pestun:2009nn} we have $\tilde A = A + i *_{2d} \Phi_t$ where $*_{2d}$ is the Hodge star\footnote{Our conventions for the Hodge star are such that in flat space with metric $ds^2=dx_1^2+dx_2^2$ we have $*_{2d} dx^1=dx^2,\,*_{2d}dx^2=-dx^1$, and the orientation on $S^2$ is the standard orientation of flat space when we use the stereographic coordinates.} on $S^2$ and $\Phi_t$ is the two-component one-form obtained from the components of the 4d field $\Phi_i$ ``tangent'' to the $S^2$, see eq. (\ref{eq:Wilson-loop}).

\subsection{Localization for local operators on $S^2$}

In \cite{Pestun:2009nn} it is shown that the 4d $\CalN=4$ SYM path-integral localizes to a 2d theory on $S^2$, namely to the constrained Hitchin/Higgs-Yang-Mills theory, or conjecturally to the zero-instanton sector of the standard bosonic Yang-Mills, which we denote as aYM theory (here ``aYM" stands for ``almost Yang-Mills", in view of the fact that contributions of the unstable instantons are dropped).

The localization computation \cite{Pestun:2009nn} implies that certain local observables inserted
on the same two-sphere where the 1/8-BPS Wilson loops are located, are also mapped to the two-dimensional 
theory. We briefly explain this fact in the following, and refer the reader to \cite{Pestun:2009nn} for more details on the localization calculation.

Choose one of the three remaining scalars which do not couple to the Wilson loops on $S^2$ and denote it $\Phi_B$. In \cite{Pestun:2009nn} it is shown that at the localization locus one has $\Phi_B = 0$, while\footnote{Literally in \cite{Pestun:2009nn} one finds the equation $d^{*2d}_A \Phi_{t} = - \Phi_n$. We changed the sign 
because of different conventions in the definition of the Wilson loop (\ref{eq:Wilson-loop})
versus \cite{Pestun:2009nn}, and also added the explicit dependence on $r$.} 
\begin{equation}
  \label{eq:local-localization}
  d^{*2d}_A \Phi_{t} = \frac {1} {r} \Phi_n,
\end{equation}
where $\Phi_{t}$ is the ``tangential" one-form obtained from $\Phi_i$ as explained above, and  $\Phi_n = \sum_{i=1}^{3} \frac {x^i}{r} \Phi_i $ denotes the component ``normal'' to the $S^2$. Here $d^{*2d}_A=*_{2d} d_A *_{2d}$ and $d_A=d+A$.
Next, in \cite{Pestun:2009nn} we find another localization equation
\begin{equation}
  \label{eq:local-localization2}
  d_A *_{2d} \Phi_{t} = i\tilde F,
\end{equation}
which relates the 4d field $\Phi_{t}$ to the 2d field $\tilde F$. Hence, combining (\ref{eq:local-localization}), (\ref{eq:local-localization2}) and $\Phi_B = 0$, we see that at the localization locus we have the relation 
\begin{equation}
\Phi_n + i \Phi_B = i r *_{2d} \tilde F.
\end{equation} 
The field $\Phi_n + i \Phi_B$ on $S^2$ is $Q$-closed, where $Q$ is the fermionic symmetry used in the localization computation (this is one of the two superconformal supersymmetries shared by the Wilson loops and local operators on $S^2$, see Section \ref{susy-sec}). Hence localization is applicable to operators which are gauge invariant functionals of this field. Therefore, the results of \cite{Pestun:2009nn} imply
that the operator $O_J(x) = \tr (\Phi_n + i \Phi_B)^J$ for $x \in S^2$ in the $\CalN=4$ 
SYM theory is mapped to the operator $ \tr ( i r *_{2d} \tilde F )^J$ in the 2d aYM, 
and hence the correlation function of any number of such operators and any number of Wilson loops are mapped to the corresponding correlation functions in the two-dimensional aYM theory.

In this paper we aim to explicitly compute some correlation functions of this type and
compare with the strong coupling limit using the dual $AdS_5 \times S^5$ description.

\subsection{Supersymmetry}
\label{susy-sec}
We now show that the supersymmetric Wilson loops on $S^2$ in (\ref{eq:Wilson-loop}) and the local operators $O_J(x) = \tr (\Phi_n + i \Phi_B)^J$ share two preserved supercharges.

It is convenient to use the notation of ${\cal N}=1$ SYM in 10d, and split the 10 Dirac matrices as $\Gamma_M=(\gamma_{\mu},\rho_A)$, where $\mu=1,\cdots,4$, $A=1,\cdots,6$, $\gamma_{\mu}$ are space-time gamma-matrices and $\rho_A$ are the $SO(6)_R$ Dirac matrices. The combined variation of the bosonic fields under Poincar\'e and superconformal supercharges can be written as 
\begin{equation}
\delta A_{\mu}=\bar\psi \gamma_{\mu}\epsilon\,,\qquad \delta \Phi_A=\bar\psi \rho_A \epsilon\,,
\end{equation} 
where $\psi$ is the gaugino and 
\begin{equation}
\epsilon=\epsilon_0+x^{\mu}\gamma_{\mu}\epsilon_1\,,
\end{equation}
where $\epsilon_0$, $\epsilon_1$ are 16-component spinors corresponding respectively to the Poincar\'e and superconformal supercharges. 

Let us first review the supersymmetries preserved by the Wilson loops (\ref{eq:Wilson-loop}). To simplify notations, we will set $r=1$ throughout this section. The loops live on $x^4=0$, $x^i x^i=1$, so we split the 4d gamma-matrices as $(\gamma_{i},\gamma_4)$. Moreover since they only couple to three of the scalars, we write $\rho_A=(\rho_i,\rho_4,\rho_5,\rho_6)$, where the index $i=1,2,3$ is identified with the space-time 3d vector index, and $\rho_4,\rho_5,\rho_6$ are rotated by $SO(3)_B \subset SO(6)_R$. The supersymmetry variation of the loop (\ref{eq:Wilson-loop}) then 
yields
\begin{equation}
\delta W \propto \dot x^j x^k \left(\gamma_{jk}\epsilon_1+i \ve_{ijk} \rho_i \epsilon_0 \right)
-x^l \gamma_l \dot x^j x^k \left(\gamma_{jk}\epsilon_0+i \ve_{ijk} \rho_i \epsilon_1 \right)\,.
\end{equation}
Therefore the variation vanishes for arbitrary loops provided that  
\begin{equation}
\gamma_{jk}\epsilon_1+i \ve_{ijk} \rho_i \epsilon_0=0\,.
\label{susy-loops}
\end{equation} 
One can eliminate for example $\epsilon_1$ from these equations to obtain the following conditions 
\begin{equation}
\left(\gamma_{ij}+\rho_{ij}\right)\epsilon_0=0\,, \quad i,j=1,2,3\,.
\label{e0-project}
\end{equation}
These are three consistent equations, but only two are independent since the commutator of any two equations gives the remaining one. With two independent projectors, we are left with 4 independent components of $\epsilon_0$. Using any of the equations in (\ref{susy-loops}), the superconformal spinor $\epsilon_1$ is completely determined in terms of $\epsilon_0$ as
\begin{equation}
\epsilon_1=-i \rho_{123}\epsilon_0\,.
\label{e1-wilson}
\end{equation} 
So the conclusion is that the loops (\ref{eq:Wilson-loop}) preserve 4 combinations of Poincar\'e and superconformal supercharges.

We now turn to the local operators on $S^2$, and let us choose $\Phi_B=\Phi_4$ to be concrete. In this case, the supersymmetry variation yields (see also \cite{Drukker:2009sf})
\begin{equation}
\delta O_J(x) \propto x^i \left(\rho_i \epsilon_0+ i \rho_4 \gamma_i \epsilon_1\right)
-x^i x^j \gamma_j \left(\rho_i \epsilon_1+i \rho_4 \gamma_i \epsilon_0\right)\,.
\end{equation}
The variation vanishes independently from the insertion point $x^i$ if 
\begin{equation}
\rho_i \epsilon_0+ i \rho_4 \gamma_i \epsilon_1=0\,.
\label{susy-local}
\end{equation}
As before, we can proceed by eliminating $\epsilon_1$ from these equations, which yields the constraints
\begin{equation}
\left(\gamma_{ij}+\rho_{ij}\right)\epsilon_0=0\,, \quad i,j=1,2,3\,,
\end{equation}
which are exactly the same conditions found for the Wilson loops above. So again we have 4 independent solutions for $\epsilon_0$. The superconformal spinor can be now determined in terms of $\epsilon_0$ from any of the (\ref{susy-local}), and the result is
\begin{equation}
\epsilon_1=-i \gamma_1 \rho_1 \rho_4 \epsilon_0\,.
\label{e1-local}
\end{equation} 
Therefore the conclusion is that the local operators on $S^2$ preserve 4 supercharges, but these are not the same 4 supercharges preserved by the Wilson loops. 

To see whether the local operators and the loops share some supercharges, we should impose that (\ref{e1-wilson}) and (\ref{e1-local}) are simultaneously satisfied. This yields the further condition on $\epsilon_0$ 
\begin{equation}
\left(\rho_{23}+\gamma_1 \rho_4\right)\epsilon_0=0\,.
\label{e0-additional}
\end{equation}
One can now see that this condition is consistent with the three equations (\ref{e0-project}), as commutators of any two of the four equations (\ref{e0-project})-(\ref{e0-additional}) either vanish or produce an equation in the same set. Therefore there are three independent projectors and hence 2 independent solutions for $\epsilon_0$. The spinor $\epsilon_1$ is given in terms of $\epsilon_0$ using either (\ref{e1-wilson}) or (\ref{e1-local}), so we conclude that the combined system of any number of local operators and any number of Wilson loops on $S^2$ preserves 2 supercharges. 

\section{Explicit perturbative checks}
\label{pert-sec}
In this section we give some explicit perturbative checks of the correspondence
$O_J(x) \leftrightarrow   \tr ( i r *_{2d} \tilde F )^J$ between 4d and 2d theory. 

One could always use conformal invariance to take the $S^2$ to have unit radius $r=1$, but we have chosen not to do so to keep dimensions of the relevant quantities in the 2d and 4d theory more transparent. For simple book-keeping let us summarize the dimensions of the relevant quantities 
in terms of unit of length $[L]$
\begin{equation}
  \label{eq:dimensions}
  \begin{aligned}
  &\phantom{}[x] = [r] =L^1 \quad 
%  [r] = L^1 \quad
  [g_{YM}^2] = L^{0} \quad
  [\Phi^i ]= L^{-1} \quad
  [F_{\mu \nu}] = L^{-2} \quad
  [W(C)] = L^{0} \quad
  [O_J] = L^{-J} \quad \\
 &[z] = L^0 \quad
  [g_{2d}^2] = L^{-2} \quad
  [\tilde F_{\bar z z} ] = L^{0} \quad
  [*_{2d} \tilde F] = L^{-2} \,,
%  [A] = [A_1] = [A_2] = L^2 
\end{aligned}
\end{equation}
where $z$ is the complex coordinate on $S^2$, see below.

In our conventions the 4d propagators for the gauge field in Feynman gauge and for the scalars are 
\begin{equation}
\la A_{\mu}^a(x) A_{\nu}^b(y)\ra =\frac{g_{YM}^2}{4\pi^2}\frac{\delta^{ab}g_{\mu\nu}}{(x-y)^2}\,,\qquad
\la \Phi_A^a(x) \Phi_B^b(y)\ra =\frac{g_{YM}^2}{4\pi^2}\frac{\delta^{ab}\delta_{A B}}{(x-y)^2}\,.
\end{equation}
In the 2d theory, we work with complex coordinates on $S^2$, with metric given by
\begin{equation}
ds^2=\frac{4 r^2 dzd\bar z}{\left(1+z\bar z\right)^2}\,,
\end{equation}
where $r$ is the radius of the sphere. This is related to the standard metric in polar coordinates by $z=\tan\frac{\theta}{2}e^{i\phi}$. 
%%%%In the following, we will set for simplicity $r=1$. 
For 2d perturbative calculations, it is convenient to use the ``Euclidean light-cone" gauge defined by\footnote{The components of the 2d gauge field $\tilde{A}_z$ and $\tilde{A}_{\bar z}$ are treated as independent. This choice of gauge is consistent for perturbative calculations, but it cannot capture the non-perturbative corrections since there are no classical solutions (instantons) that satisfy this gauge.}
\begin{equation}
\tilde A_{\bar z}=0\,.
\end{equation}
In this gauge there are no interactions and the 2d YM action becomes simply
\begin{equation}
\label{eq:2daction-light-cone}
S_{YM_2}=-\frac{1}{2g_{2d}^2}\int d^2z \sqrt{g} g^{z\bar z}\partial_{\bar z}\tilde{A}_z^a g^{z\bar z}\partial_{\bar z} \tilde{A}_z^a\,.
\end{equation}
We use notations $d^2 z = d \bar z \wedge dz = 2 i dx \wedge dy$ for $z = x + i y$, and $\sqrt{g} = -i g_{\bar z z}$,
so that $d^2 z \sqrt{g}$ is the conventional volume form on $S^2$ normalized as 
\begin{equation}
  \label{eq:normalization}
  \int {d^2 z \sqrt{g}} = 4 \pi r^2.
\end{equation}
The gauge field propagator is \cite{Drukker:2007yx}\footnote{In \cite{Drukker:2007yx} the gauge field propagator is given in a certain generalized Feynman gauge, in which both $\tilde{A}_{z}$ and $\tilde{A}_{\bar z}$ are propagating, and the $\tilde{A}_z$ propagator is related to the one in the ``light-cone" gauge used here by a factor of 2.} (see also the Appendix)
\begin{equation}
\la \tilde{A}_z^a(z) \tilde{A}_z^b(w)\ra = \frac{g^2_{2d} r^2 }{\pi} \delta^{ab} \frac{1}{1+z\bar z}\frac{1}{1+w \bar w}\frac{\bar z-\bar w}{z-w}\,.
\label{light-prop}
\end{equation} 
This satisfies\footnote{Our convention is such that $\int d^2 z \delta^2(z-w) f(z,\bar z )  = f(w,\bar w)$ for any $f(w,\bar w)$.}
\begin{equation}
\frac{1}{g^2_{2d}}\partial_{\bar z} \left(\sqrt{g}(g^{z\bar z})^2 \partial_{\bar z} \la \tilde{A}_z^a(z) \tilde{A}_z^b(w) \ra\right) = \delta^{ab}\delta^2(z-w)\,.
\end{equation}

On $S^2$ there is no ambiguity in the propagator for the kinetic
term (\ref{eq:2daction-light-cone}). A quick explanation
is that the kinetic term operator is the square of
 the Dolbeault operator $\bar \p$,
which maps $(1,0)$-forms on $S^2 \simeq \CP^1$, 
which are sections of the bundle $\CalO(-2)$, 
to the $(1,1)$-forms. 
However, the bundle $\CalO(-2)$ does not have holomorphic sections,
hence there are no zero modes, and the propagator is well defined. 

We can also explicitly show that there is no ambiguity by solving the equation 
\newcommand{ \pbar }{\p_{\bar z}}
\begin{equation}
  \label{eq:D-equation}
  \pbar (\rho^{-2} \pbar G_z (z,\bar z)) =  0
\end{equation}
on $\BC$ and analyzing the behaviour at infinity. Here we denoted $\rho = (1+ z \bar z )^{-1}$.
From (\ref{eq:D-equation}) we get
\begin{equation}
  \label{eq:eq-D-equation2}
  \pbar G_z (z,\bar z) = \rho^2 f(z)\,,
\end{equation}
where $f(z)$ is an arbitrary holomorphic function on $\BC$.
Solving (\ref{eq:eq-D-equation2}), we get
\begin{equation}
  \label{eq:eq-D-equation3}
  G_z (z, \bar z) = \frac {\bar z} { 1 + z \bar z} f(z) + g(z)\,,
\end{equation}
where $g(z)$ is again an arbitrary holomorphic function on $\BC$. Now we must require that 
$G_z$ at $z \to \infty$ decreases at least as fast as $z^{-2}$ in order 
for the solution to be smooth on $\CP^{1}$. Indeed, if we make a coordinate
transformation to the coordinate $\tilde z =1/z$ so that the point $z=\infty$ maps 
to the point $\tilde z =0$, we get $G_z dz = G_{\tilde z} d\tilde z  = - G_{\tilde z} \frac { dz } { z^2}$.
Asking $G_{\tilde z}$ to be finite at $\tilde z=0$,
 we see that $G_z$ decreases 
at least as $z^{-2}$ at $z\to \infty$. However, the solution (\ref{eq:eq-D-equation3}) implies
that $G_z (z,\bar z)$  decreases not faster than $z^{-1}$, 
unless both $f(z)$ and $g(z)$ vanish. Therefore, the equation (\ref{eq:D-equation}) does not have non-zero smooth solutions on $\CP^1$. The actual solution (\ref{light-prop}) has the correct asymptotics $z^{-2}$ at $z \to \infty$, hence it exists and is well defined globally on  $\CP^{1}$. 

For later convenience, we also write down the explicit expression for the scalar dual to the 2d field strength in these coordinates
\begin{equation}
*_{2d} \tilde F = \frac 1 {\sqrt{g}}\tilde{F}_{\bar z z}= \frac{i}{2 r^2} \left(1+z \bar z\right)^2 \partial_{\bar z}\tilde{A}_{z}\,.
\label{dual-scalar}
\end{equation}

\subsection{Correlators of local operators} 
In the 4d theory, the tree level 2-point function of the elementary fields making up the local operators $O_J(x)$ on $S^2$ is (here and in the following we pick $\Phi_B=\Phi_4$ to be concrete)
\begin{equation}
\begin{aligned}
\langle (\Phi_n + i \Phi_4)^a(x) (\Phi_n + i \Phi_4)^b(y) \rangle &= \frac 1
{r^2} x^i y^j \langle \Phi_i^a(x) \Phi_j^b(y) \rangle-\langle \Phi_4^a(x) \Phi_4^b(y)\rangle\\
&=\frac{g_{YM}^2}{4\pi^2}\delta^{ab} \left( \frac{x\cdot
    y/r^2-1}{(x-y)^2}\right)=-\frac{g_{YM}^2}{8\pi^2 r^2}\delta^{ab}\,,
\label{propag-OJ}
\end{aligned}
\end{equation}
where we have used $x^2=y^2=r^2$ since the operators are inserted on $S^2$.
Thus we see that correlation functions between the local operators $O_J(x)=\tr (\Phi_n + i \Phi_4)^J$ are position independent at tree level. This was also observed in \cite{Drukker:2009sf} (for the more general operators on $\mathbb{R}^4$ of which our operators on $S^2$ are a special case), and it was argued that position independence holds true even at the quantum level by using a Ward identity which follows from the preserved supersymmetries. It was further argued in \cite{Drukker:2009sf}, based on the results of \cite{deMedeiros:2001kx}, that correlation functions involving an arbitrary number of $O_J(x)$ do not receive quantum corrections (up to possible instanton corrections) and thus can be obtained by simply doing Wick contractions with the free propagator (\ref{propag-OJ}). The position independence implies that the $n$-point function can be computed by a Gaussian multi-matrix model. It is not difficult to see that the following  $n$-matrix model correctly captures the gauge group combinatorics
\begin{equation}
Z_{nMM}=\int [dX_1^{a}]\cdots[dX_n^{a}]\,e^{\frac{4 \pi^2 r^2}{g_{YM}^2}(\frac {1}{n-1} (\sum_{i} X^a_i)^2  - 
\sum_{i} (X^a_i)^2  )}\,,
\label{n-MM}
\end{equation}
where $X_k=X_k^a T_a$ are matrices in the Lie algebra of the gauge group. Due to position independence, the tree-level $n$-point functions are then given by
\begin{equation}
\la O_{J_1}(x_1)O_{J_2}(x_2)\cdots O_{J_n}(x_n) \ra=\la \tr X_1^{J_1} \tr X_2^{J_2}\cdots \tr X_n^{J_n}\ra_{nMM}\,,
\end{equation}
where the correlation function on the right-hand side is taken in the $n$-matrix model (\ref{n-MM}). 
Inverting the quadratic form in the matrix model action, one can see that there are only propagators
between different matrices, as appropriate for normal-ordered operators $O_J(x)$, and the normalization of the action in (\ref{n-MM}) is chosen to match the propagator (\ref{propag-OJ}). 

Let us now turn to the 2d theory. To check the correspondence between $O_J(x)$
and $\tr(i r *_{2d} \tilde F)^J$, let us compute the 2-point function
\begin{equation}
\langle i r *_{2d} \tilde F^a(z)\, i r *_{2d} \tilde F^b(w)\rangle = \frac{1}{4 r^2}(1+z \bar z)^2 (1+w \bar w)^2 \langle \partial_{\bar z} \tilde{A}^a_z(z) \partial_{\bar w} \tilde{A}^b_z(w)\rangle\,.
\end{equation}
Using the light-cone gauge propagator (\ref{light-prop}) we get
\begin{equation}
\langle i r *_{2d} \tilde F^a(z)\, i r *_{2d} \tilde F^b(w)\rangle
=\delta^{ab}\left( \frac{g_{2d}^2 }{4\pi }-
\frac{i g^2_{2d} }{2}(1+z \bar z)^2 \delta^2(z-w)\right)\,.
\label{prop-2d}
\end{equation}

The $\delta$-function piece does not matter if we consider correlation functions
of normal ordered operators inserted at distinct points. If we ignore the
$\delta$-function term, we then see that this propagator agrees with
(\ref{propag-OJ}) provided we identify the 2d and 4d couplings as
$g_{2d}^2=-\frac{g_{YM}^2}{2\pi r^2}$, as implied by the conjecture of
\cite{Drukker:2007yx,Drukker:2007qr} for Wilson loops on $S^2$ and by the
localization calculation of \cite{Pestun:2009nn}. Since in the light-cone
gauge the 2d YM action is free, correlation functions of $\tr (i *_{2d} \tilde
F(z))^J$ inserted at arbitrary points will be also given, perturbatively, by
just doing contractions with the free propagator (\ref{prop-2d}). The
combinatorics for the contraction of gauge indices is the same as the one for
the 4d local operators, so we can conclude that 
\begin{equation}
\langle O_{J_1}(x_1)\cdots O_{J_n}(x_n)\rangle_{4d} = 
\langle \tr\left(i r *_{2d} \tilde F(z_1)\right)^{J_1}\cdots \tr \left(i r *_{2d} \tilde F(z_n)\right)^{J_n}\rangle_{2d}^{\mbox{\tiny{0-inst}}}\,,
\label{npoint-2d4d}
\end{equation}   
where the equality holds for normal-ordered operators inserted at distinct points. The non-renormalization arguments of \cite{Drukker:2009sf} and the localization argument we propose in this paper suggest that the relation (\ref{npoint-2d4d}) should hold to all orders in perturbation theory (possible instanton contributions on the 4d side are not in principle excluded). 

\subsection{Correlator of a local operator and a Wilson loop}
\begin{figure}
\begin{center}
\includegraphics[width=40mm]{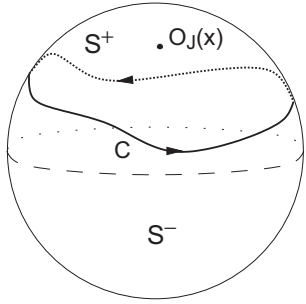}
\parbox{13cm}{\caption{The correlator of a Wilson loop $W(C)$ and a local operator $O_J(x)$ on $S^2$. The curve $C$ divides the $S^2$ into two regions which we denote as $S^+$ and $S^-$. Perturbatively, we find that the correlator only depends on whether the operator is inserted in the $S^+$ or $S^-$ region.}
\label{OJW}}
\end{center}
\end{figure}

We now consider the correlation function of a local operator and a Wilson loop on $S^2$, to leading order in perturbation theory. We will start by considering the correlation function involving a single elementary field $\Phi_n+i \Phi_4$, and let us suppress the gauge indices for the moment. In the 4d theory, we wish to compute
\begin{equation}
\langle \left(\frac {x^l}{r} \Phi_l(x) + i \Phi_4(x)\right) W(C)  \rangle\,.
\end{equation}
Expanding the path-ordered exponential in the Wilson loop, to leading order we have to evaluate
\begin{equation}
\langle \left(\frac {x^l}{r} \Phi_l(x) + i \Phi_4(x)\right)   
\oint_C (A_j  + i \ve_{ijk} \Phi_i \frac {y^k}{r} ) dy^j\rangle=i
\frac{g^2_{YM}}{4\pi^2 r^2 } \oint_C dy^j \frac{\ve_{ijk} x^i y^k}{(x-y)^2}\,.
\label{local-to-loop-4d}
\end{equation}
It is not difficult to compute this integral for arbitrary loop, see also \cite{Drukker:2007yx}. Define $\theta$ to be the angle between $x^i$ and $y^i$. If we denote by $d\phi$ the one-form orthogonal to $d\theta$, then we get
\begin{equation}
\oint_C dy^j \frac{\ve_{ijk} x^i y^k}{(x-y)^2} = -\oint_C d\phi \frac{\sin^2\theta}{2(1-\cos\theta)}=
 -\oint_C d\phi \cos^2\frac{\theta}{2}\,.
\end{equation}
For $r=1$ this is equal to half the area of the region of $S^2$ enclosed by the loop and
not containing the point $x$, up to a choice of $+/-$ sign which depends on the orientation of the loop relative to the point $x$. If we denote by $S^+$,$S^-$ the two regions of $S^2$ singled out by the loop, see Figure \ref{OJW}, such that $S^+$ is the one containing the north pole, and if we take the loop to run counterclockwise with respect to the north pole, then we can summarize the result as 
\begin{equation}
\langle \left(\frac {x^l}{r} \Phi_l(x) + i \Phi_4(x)\right) \oint_C (A_j  + i \ve_{ijk} \Phi_i \frac {y^k}{r} ) dy^j  \rangle
= \begin{cases}  -i \frac{g_{YM}^2}{2\pi r}
  \frac{A_2}{A} \,,\quad x\in S^+\\
 +i \frac {g_{YM}^2}{2 \pi r} \frac {A_1}{A}
\,,\quad x\in S^- \end{cases}\,,
%+\,{\cal O}(g_{YM}^4)
%-i \frac{g_{YM}^2}{8\pi^2 r^3} A_2 =
%+i \frac{g_{YM}^2}{8\pi^2 r^3} A_1 =
\label{OW-4d}
\end{equation}
where $A_1$ and $A_2$ are the areas of $S^+$ and $S^-$ respectively. It can be seen that this is precisely the behavior expected in the 2d Yang-Mills theory, in particular we see that the result only depends on the area and not on the shape of the loop. Moreover, it is almost independent from the position of the local operator, it only depends on whether the operator is inserted ``inside" or ``outside" the loop. 

Let us now carry out the analogous calculation in the 2d theory. Here, to first order in the coupling, we have to compute 
\begin{equation}
\langle i r *_{2d} \tilde{F}(z) \, \oint_C dw \tilde{A}_w \rangle\,. 
\end{equation}
Since we are working at first order in perturbation theory, we can essentially assume that we are in the abelian theory. Then by Stokes' theorem we can write
\begin{equation}
\oint_C \tilde{A} = \int_{S^+} \tilde{F}\,,
\end{equation}
where $S^+$ is the region enclosed by the loop and containing the origin of the complex plane (i.e. the north pole). Using (\ref{dual-scalar}) we get
\begin{equation}
\begin{aligned}
\langle i r *_{2d} \tilde{F}(z) \, \oint_C dw \tilde{A}_w \rangle &= -
\int_{S^+}d^2w \frac{\,2 r}{(1+w \bar w)^2} \langle i r *_{2d} \tilde F(z)\, i r *_{2d} \tilde F(w)\rangle \\
&= - \frac{ g^2_{2d} r }{4\pi}\int_{S^+} d^2w \frac{\,2}{(1+w \bar w)^2}
\left(1-2\pi i (1+w \bar w)^2\delta^2(z-w)\right)\\
&= \begin{cases} +i \frac{g_{2d}^2}{4\pi r} A_2\,,\quad z\in S^+\\
-i \frac{g_{2d}^2}{4\pi r} A_1\,,\quad z\in S^-\,, \end{cases}
\end{aligned}
\label{FW-2d}
\end{equation}
which indeed agrees with the 4d result (\ref{OW-4d}) by virtue of the
identification $g_{2d}^2=-g_{YM}^2/2\pi r^2$. As a double-check of factors of $i$ and signs, let us also do this computation in a different way using Cauchy's theorem. Using the invariance under area preserving diffeomorphisms of 2d YM, we can for simplicity consider a latitude on $S^2$, which is a circle of radius $\tan\frac{\theta_0}{2}$ in the $z,\bar z$ coordinates (here $\theta_0$ is the latitude angle). Then
\begin{equation}
\begin{aligned}
\langle i r *_{2d} \tilde{F}(z) \, \oint_C dw \tilde{A}_w \rangle &=-\frac{1}{2 r}(1+z \bar z)^2 \oint_{|w|=\tan\frac{\theta_0}{2}}\!\!\!\!\!\!\!\!dw\,
\langle \partial_{\bar z} A_z(z) A_z(w)  \rangle \\
&=\frac{g_{2d}^2 r}{2\pi} \cos^2\frac{\theta_0}{2}\oint_{|w|=\tan\frac{\theta_0}{2}}\!\!\!\!\!\!\!\!dw\,
\left(\frac{1}{w-z}+\tan^2\frac{\theta_0}{2}\,\frac{z}{w(w-z)}\right)\\
&=\begin{cases}+i \frac{g^2_{2d} r}{4\pi} \cdot 4\pi
  \cos^2\frac{\theta_0}{2}\,,\quad |z|<\tan\frac{\theta_0}{2}\\-i \frac{g^2_{2d}
  r }{4\pi} \cdot 4\pi \sin^2\frac{\theta_0}{2}\,,\quad |z|>\tan\frac{\theta_0}{2}\,,\end{cases}
\end{aligned}
\label{Cauchy-latit}
\end{equation}
where we have used that $\bar w = r_0^2/w$ for a circular contour of radius $r_0$, and in the last equality we have used Cauchy's theorem. As expected this
agrees with (\ref{FW-2d}), since $A_2=4\pi r^2 \cos^2\frac{\theta_0}{2}$ and
$A_1=4\pi r^2 \sin^2 \frac{\theta_0}{2}$ for a latitude circle.

It is straightforward to extend the above calculations to the case of the
 correlator $\la O_J(x)W(C)\ra_{4d}$ (or equivalently $\la \tr (i r
 *_{2d}\tilde{F})^J W(C)\ra_{2d}$) for arbitrary $J$. To first order in
 perturbation theory, we consider a Feynman diagram obtained by joining the
 local operator to the Wilson loop with $J$ ``local-to-loop" propagators as
 computed in eq. (\ref{local-to-loop-4d}) (or eq. (\ref{FW-2d})). Since the
 product of the $J$-propagators is symmetric in the exchange of the vertices,
 path ordering of the exponential in the Wilson loop does not
 matter. Considering only the planar contractions we then get
\begin{equation}
\begin{aligned}
\la O_J(x) W(C) \ra&= \frac{1}{N}\left(-\frac{N}{2}\right)^J \frac{J}{J!} \left( i \frac{g^2_{YM}}{4\pi^2} \oint_C dy^j \frac{\ve_{ijk} x^i y^k}{(x-y)^2}\right)^J + {\cal O}(g^{2J+1}_{YM})\\
&=\frac{i^J}{N} \frac{\lambda^J}{(J-1)!}\begin{cases}\left(\frac{A_2 r}{A^2}\right)^J\,,\quad x\in S^+ \\
 \left(\frac{-A_1 r}{A^2}\right)^J\,,\quad x\in S^- 
\end{cases}+{\cal O}(\lambda^{J+1})\,,
\end{aligned}
\label{weak-coupling}
\end{equation}
where $\lambda=g_{YM}^2 N$ is the `t Hooft coupling. In the first line the factor $1/N$ comes from the normalization of the Wilson loop (we take $R$ to be the fundamental representation), the factor $(-N/2)^J$ comes from the contractions of gauge group generators, and the factor of $J$ counts the number of planar diagrams obtained from cyclic permutations. At the next orders in perturbation theory, one decorates the above Feynman diagram by ladder ``loop-to-loop" propagators, as well as internal interaction vertices and loops in the case of the 4d theory. In the 2d YM theory in light-cone gauge, on the other hand, there are no interactions and the full perturbative result is given by summing up ladder diagrams. If the conjectured 4d-2d correspondence is correct, the sum of the light-cone ladder diagrams in the 2d theory should be equal to the sum of all Feynman diagrams (including interacting diagrams) in the 4d SYM theory. In the next section we will write down a Gaussian two-matrix model which computes the sum of the 2d light-cone ladder diagrams. 

\section{A matrix model for the correlator of $O_J$ and $W_{R}(C)$}
\label{2mm-sec}
Let us start by considering for simplicity the case of a latitude on $S^2$. The result can then be generalized to arbitrary loops by using the invariance under area preserving diffeomorphisms of the 2d YM theory. In the case of the latitude, it is especially simple to sum up ladder diagrams, because the ``loop-to-loop'' propagator is a constant. Let us parameterize the loop as $z(\tau)=r_0 e^{i \tau}$, where $r_0=\tan \frac{\theta_0}{2}$ and $\theta_0$ is the latitude angle. Then the propagator between two points $\tau_1$,$\tau_2$ along the loop is, using (\ref{light-prop}), 
\begin{equation}
\la \dot z_1 A^a_z(z_1) \dot z_2 A^b_z(z_2)  \ra = \delta^{ab} \frac{g_{2d}^2 r^2}{\pi}\frac{r_0^2}{(1+r_0^2)^2}=\delta^{ab} \frac{g^2_{2d}}{(2\pi)^2}\frac{A_1 A_2}{A}\,,
\label{loop-to-loop}  
\end{equation}
where in the last equality we have written the result in terms of the areas singled out by the loop, and $A=A_1+A_2$. Since the loop-to-loop propagator is a constant, path ordering is not important and the calculation of the previous section directly shows that the correlator is independent of the insertion point of the local operator. Hence, specializing to the case in which the local operator sits in $S^+$, we can effectively consider the insertion point to be the north pole, i.e. $z=0$. Then the local-to-loop propagator is also a constant, see eq. (\ref{Cauchy-latit})
\begin{equation}
\la i r  *_{2d} F^a(0)\, \dot z A^b_z(z) \ra = i\delta^{ab} \frac{g_{2d}^2
  r}{2\pi}\frac{1}{1+r_0^2}= i\delta^{ab} \frac{g_{2d}^2 r}{2\pi}\frac{A_2}{A}\,.
\label{local-to-loop}
\end{equation}
This allows us to easily sum up the 2d light-cone ladder diagrams in terms of the following two-matrix model, which we write directly in terms of the 4d coupling using the relation $g^2_{2d}=-\frac{2}{A}g^2_{YM}$
\begin{equation}
Z_{2MM}=\int [dX][dY]\, e^{-\frac{A^2}{2g_{YM}^2}\left(\frac{A_1}{A_2 r^2}\tr
    Y^2-\frac{2i}{A_2 r}\tr X Y\right)}\,,
\label{2MM}
\end{equation}
where $X$ and $Y$ are $N\times N$ Hermitian matrices. The propagators in this matrix model are
\begin{equation}
\la X^i_j X^k_l \ra = g_{YM}^2 \frac{A_1 A_2}{A^2} \delta^i_l \delta^k_j \,,\qquad 
\la Y^i_j X^k_l \ra = i g_{YM}^2 \frac{r A_2}{A^2} \delta^i_l \delta^k_j \,,
\end{equation}
which match respectively the loop-to-loop and the local-to-loop propagators,
eq. (\ref{loop-to-loop}) and (\ref{local-to-loop}) (to make the comparison, use
$g^2_{2d}=-\frac{2}{A}g^2_{YM}$ and recall that in our conventions the $U(N)$
generators satisfy $(T^a)^i_j (T^a)^k_l = -\frac{1}{2} \delta^i_l
\delta^k_j$, and that we integrate over the loop $\tau \in [0,2\pi]$). The $\la Y Y\ra$ propagator vanishes as required by normal
ordering of the local operator. Therefore the correlators of interest are given
by the two-matrix model correlation functions 
\begin{equation}
\la \tr (i r *_{2d} \tilde F(z))^J\,W_R(C) \ra_{2d}^{\mbox{\tiny{0-inst}}}=\la \tr Y^J \frac{1}{\dimR} \tr_R e^X \ra_{2MM}\,,
\end{equation}
where the label ``0-inst" is to remind us that the sum of the 2d light-cone
ladder diagrams does not capture the non-perturbative contributions of the
unstable instantons on $S^2$. In the above we have assumed that the local
operator is inserted in the region including the north pole, and we will do the same in what 
follows. If the operator is inserted in the complementary region, one simply exchanges the roles of $A_1$ and $A_2$ and includes an additional factor of $(-1)^J$ due to orientation. 

 The arguments of \cite{Drukker:2007qr,Drukker:2007yx}, the localization
 calculation of \cite{Pestun:2009nn} and the identification $\tr (i r *_{2d}
 \tilde F)^J \leftrightarrow O_J(x)$ lead us then to propose that the same
 two-matrix model may exactly capture the correlators in the 4d ${\cal N}=4$ SYM
 theory 
\begin{equation}
\la O_J(x) W_R(C) \ra_{4d} = \frac{1}{Z_{2MM}} \int [dX][dY]\, \tr Y^J \frac{1}{\dimR}\tr_R e^X\,
e^{-\frac{A^2}{2g_{YM}^2}\left(\frac{A_1}{r^2 A_2}\tr Y^2-\frac{2i}{r A_2}\tr X Y\right)}\,.
\label{2MM-4d}
\end{equation}
Note that for general loops this matrix model is not equal to the sum of the ladder diagrams in the 4d theory, since the combined gauge-scalar ladder propagator is not equal to the 2d light-cone propagator, but rather to its real part \cite{Drukker:2007qr,Drukker:2007yx}. If the loop-to-loop propagator is a constant, which happens in the case of the 1/4 BPS latitude, then the matrix model is indeed equal to the sum of 4d ladder diagrams, and the proposed relation to 2d YM would imply that the sum of interacting diagrams should vanish. It may be possible to explicitly check this, to leading order in perturbation theory, by adapting to our case the results of \cite{Erickson:2000af,Semenoff:2001xp,Pestun:2002mr,Drukker:2006ga,Semenoff:2006am}. For more general loops, on the other hand, one should check whether the sum of ladder and interacting diagrams in 4d is equal to the 2d light-cone ladder diagrams. In the case of the 1/4 BPS loop made up of two arcs of longitudes \cite{Drukker:2007qr}, this has been successfully checked in \cite{Bassetto:2008yf,Young:2008ed} for the case of the expectation value of the Wilson loop. It should be possible to extend those calculations to the case of the correlation function with a local operator.

As a consistency check of the two-matrix model, notice that if we do not insert the local operator, then we can integrate out $Y$ exactly and we end up with
\begin{equation}
\la W_R(C) \ra_{4d} = \frac{1}{Z_{MM}} \int [dX] \frac{1}{\dimR}\tr_R e^X\, 
e^{-\frac{A^2}{2 A_1 A_2 g_{YM}^2}\tr X^2}\,,
\end{equation}
which is precisely the matrix model proposed in \cite{Drukker:2007qr,Drukker:2007yx} to exactly capture the expectation value of the 1/8 BPS Wilson loops on $S^2$. In the case of a loop at the equator, $A_1=A_2=A/2$, it reduces to the well-known matrix model for the 1/2 BPS circular loop \cite{Erickson:2000af,Drukker:2000rr}.

In the next subsection we will solve the two-matrix model in the planar limit. Remarkably, the strong coupling behavior precisely agrees with the dual string theory calculation we present in Section \ref{string-sec}.

\subsection{Large $N$ solution}
It is straightforward to solve the two-matrix model (\ref{2MM-4d}) in the large $N$ limit, following \cite{eynard-1997} (see also \cite{Giombi:2009ms} for more details on the application of the formalism of \cite{eynard-1997} to the Gaussian two-matrix model). Since the action is Gaussian, it is also straightforward to solve the matrix model at finite $N$, but we do not do this here. For simplicity, in the following we will restrict to the case of a Wilson loop in the fundamental representation.

For a general Gaussian two-matrix model with action
\begin{equation}
S_{2MM}= N\left(\frac{a_1}{2}\tr X^2+\frac{a_2}{2}\tr Y^2-c \tr X Y \right)\,,
\end{equation}
the planar two-point resolvent 
\begin{equation}
\omega(z_1,z_2)=\la \tr \frac{1}{z_1-X} \tr \frac{1}{z_2-Y} \ra_{\mbox{\tiny conn}}\,,
\label{resolv}
\end{equation}
is given by 
\begin{equation}
\omega(z_1,z_2)=-\partial_{z_1}\partial_{z_2} \log \left(1-y_1(z_1) y_2(z_2)\right)\,,
\label{resolv2}
\end{equation}
where $y_1(z_1)$ and $y_2(z_2)$ are two resolvent functions
\begin{equation}
y_1(z_1)=\frac{1}{2\alpha}\left(z_1+\sqrt{z_1^2-4\alpha \alpha_1}\right)\,,\quad 
y_2(z_2)=\frac{1}{2\alpha}\left(z_2+\sqrt{z_2^2-4\alpha \alpha_2}\right)\,.
\end{equation}
Here the parameters $\alpha$, $\alpha_1$ and $\alpha_2$ are related to the parameters in the action by
\begin{equation}
\alpha^2=\frac{c}{a_1 a_2-c^2}\,,\quad \alpha_1 = \frac{a_2}{c}\alpha \,,\quad 
\alpha_2=\frac{a_1}{c}\alpha \,. 
\end{equation}
For the two-matrix model (\ref{2MM}), we get
\begin{equation}
\alpha^2=i \lambda \frac{A_2}{A^2} r\,,\quad 4\alpha \alpha_1=\frac{4 A_1 A_2}{A^2} \lambda\,,\quad \alpha_2=0\,.
\end{equation}
As it is clear from the definition (\ref{resolv}), to extract the correlators $\la \tr Y^J \tr e^X \ra$ we should expand the resolvent in inverse powers of $z_2$
\begin{equation}
\omega(z_1,z_2)=\sum_{n=1}^{\infty} \omega_n(z_1) z_2^{-n-1}\,,
\end{equation}
extract the term with $n=J$, and perform the inverse Laplace transform on the $z_1$ variable
\begin{equation}
W_J(t_1)=\frac{1}{2\pi i} \oint dz_1 \omega_J(z_1) e^{t_1 z_1}\,.
\end{equation}
Then
\begin{equation}
\la \tr Y^J \tr e^{t_1 X} \ra = W_J(t_1)\,.
\end{equation}
By expanding (\ref{resolv2}), we get
\begin{equation}
\omega_J(z_1) = J \alpha^J y_1^{-J-1} \partial_{z_1} y_1\,.
\end{equation}  
Using the explicit expression for $y_1(z_1)$, the inverse Laplace transform yields
\begin{equation}
W_J(t_1)= J \alpha^J \left(\frac{\alpha}{\alpha_1}\right)^{\frac{J}{2}}\, I_J(2 t_1 \sqrt{\alpha \alpha_1})\,,
\end{equation}
where $I_J(x)$ is a modified Bessel function of the first kind
\begin{equation}
I_J(x)=\frac{1}{2\pi} \int_0^{2\pi} d\phi\, e^{i J \phi} e^{x \cos\phi}\,.
\end{equation}
Putting everything together, our final result for the correlator in the planar limit is then 
\begin{equation}
\la O_J(x)W(C)\ra=\la \tr Y^J \frac{1}{N}\tr e^X \ra = \frac{i^J}{N} \frac{J
  \lambda^{J/2} r^J}{A^J} 
\left(\frac{A_2}{A_1}\right)^{\frac{J}{2}} I_J\left(\sqrt{\frac{4 A_1 A_2}{A^2}\lambda}\right)\,.
\label{final-2MM}
\end{equation}
Expanding this expression at weak coupling, we get
\begin{equation}
\la O_J(x)W(C)\ra = \frac{i^J}{N} \frac{\lambda^J}{(J-1)!} \left(\frac{A_2 r}{A^2}\right)^J\left(1+ \frac{A_1A_2}{A^2} \frac{\lambda}{J+1}+{\cal O}(\lambda^2)\right)\,,
\end{equation}
which agrees, as expected, with the leading order perturbative result (\ref{weak-coupling}).

Let us now extract the strong coupling asymptotics of (\ref{final-2MM}). To compare to the string theory calculation, it is somewhat more convenient to work with the normalized local operators, see Section \ref{string-sec}
\begin{equation}
\tilde{O}_J(x) =2^{J/2} (-i)^J  \frac{(2\pi)^J}{\lambda^{J/2} \sqrt{J}} O_J(x)\,.
\end{equation}
At strong coupling it is also natural to normalize the correlator by the Wilson loop expectation value, which is given by \cite{Erickson:2000af,Drukker:2007qr,Drukker:2007yx}
\begin{equation}
\la W(C) \ra = \frac{2}{\sqrt{\lambda '}} I_1(\sqrt{\lambda '})\,,\qquad 
\lambda ' = \frac{4 A_1 A_2}{A^2}\lambda\,.
\end{equation}
The normalized correlator then reads
\begin{equation}
\frac{\la \tilde{O}_J(x) W(C)\ra}{\la W(C)\ra}=\frac {r^{-J}}{N} 2^{-J/2} \sqrt{J \lambda} \left(\frac{A_2}{A}\right)^{\frac{J}{2}+\frac{1}{2}} \left(\frac{A_1}{A}\right)^{-\frac{J}{2}+\frac{1}{2}} \frac{I_J(\sqrt{\lambda '})}{I_1(\sqrt{\lambda '})}\,.
\end{equation}
Using the asymptotic expansion\footnote{Here we only include the contribution of the dominant saddle point at $x \rightarrow \infty$.}
\begin{equation}
\frac{I_J(x)}{I_1(x)}=1-\frac{J^2-1}{2x}+\frac{J^4-4J^2+3}{8x^2}+\ldots\,,
\end{equation}
we get the following large $\lambda$ expansion of the correlator
\begin{equation}
\begin{aligned}
\frac{\la \tilde{O}_J(x) W(C)\ra}{\la W(C)\ra}= \frac{r^{-J}}{N} 2^{-J/2} \sqrt{J \lambda} \left(\frac{A_2}{A}\right)^{\frac{J}{2}+\frac{1}{2}} \left(\frac{A_1}{A}\right)^{-\frac{J}{2}+\frac{1}{2}} \left(1-\frac{J^2-1}{2\sqrt{\lambda '}}+{\cal O}(\frac{1}{\lambda})\right)\,.
\end{aligned}
\label{2MM-strong}
\end{equation}
The leading term is in precise agreement with the dual string theory on $AdS_5\times S^5$, as shown in Section \ref{string-sec}.

\subsection{Equivalence with the complex matrix model}

The two-matrix model (\ref{2MM-4d}) can be written in complex notations, so that one can see the equivalence with the complex matrix model which was first proposed in \cite{Semenoff:2001xp} to compute the correlator of a circular 1/2 BPS Wilson loop and a chiral primary operator, see \cite{Semenoff:2002kk,Zarembo:2002ph,Pestun:2002mr,Okuyama:2006jc,Giombi:2006de,Gomis:2008qa} for related work and \cite{Semenoff:2006am} for the extension to the circular 1/4 BPS Wilson loop. 

First we rewrite the matrix model (\ref{2MM-4d}) as
\begin{equation}
  \label{eq:2MM-rewrite}
  \la O_J(x) W_R(C) \ra_{4d} = \frac{1}{Z_{2MM}} \int [dX][dY]\, \tr Y^J \frac{1}{\dimR} \tr_R e^X \,
%e^{-\frac{A^2}{2g_{YM}^2}\left(\frac{A_1}{r^2 A_2}\tr Y^2-\frac{2i}{r A_2}\tr X Y\right)}\,.
e^{-\frac {A^2}{2 g_{YM}^2 A_1 A_2} \tr \left (  \frac {-iA_1}{r} Y (2 X + \frac {i A_1}{r} Y)  \right)}.
\end{equation}
Next we introduce new complex matrices
\begin{equation}
  \label{eq:new-variables}
  Z = \frac {-i A_1}{r} Y \quad 
  \bar Z =  2X + \frac {iA_1}{r} Y
\end{equation}
and change variables in the matrix integral (\ref{eq:2MM-rewrite}) from $X$ and $Y$ to $Z$ and $\bar Z$.\footnote{Note that $Z$ and $\bar Z$ defined in (\ref{eq:new-variables}) are not complex conjugate to each other, but the number of degrees of freedom is still the same. Perturbatively, we can always rotate formally the integration contour so that $Z$ and $\bar Z$ are complex conjugates.} 
The Jacobian of this change of variables is a  number independent of $X$ and $Y$,
which only changes the overall normalization of 
the partition function. Since we normalize the correlation functions in the matrix model by the partition 
function, this Jacobian does not matter. Therefore we get 
\begin{equation}
  \label{eq:2MM-rewriteZZ}
  \la O_J(x) W_R(C) \ra_{4d} = \left(\frac {i r}{A_1} \right)^J \frac{1}{Z_{Z\bar Z-model}} \int [dZ \, d\bar Z] \,
 \tr Z^J \frac{ 1 }{\dimR} \tr_R e^{\frac 1 2(Z+\bar Z)}\,
e^{-\frac {A^2}{2 g_{YM}^2 A_1 A_2} \tr Z \bar Z}.
\end{equation}
This agrees with the complex matrix model of \cite{Semenoff:2001xp} and its extension to the 1/4 BPS circular loop \cite{Semenoff:2006am}. The $i$ in the factor $(ir/A_1)^J$  takes into account that $\Phi$ in our conventions is valued in the Lie algebra and thus is an anti-Hermitian matrix for the $\SU(N)$ theory, and the $r/A_1$ comes from the finite-distance propagator between our local operator and the Wilson loop. The coefficient in the action of the complex matrix model in the case of a latitude at polar angle $\theta$ reduces to the familiar~\cite{Semenoff:2006am}
\begin{equation}
  \frac {A^2}{2 g_{YM}^2 A_1 A_2} \tr Z \bar Z = \frac { 2 }{g_{YM}^2 \sin^2 \theta} \tr Z \bar Z.
\end{equation}

Note that in earlier works %\cite{Semenoff:2001xp,Pestun:2002mr,Semenoff:2002kk,Semenoff:2006am,Okuyama:2006jc,Giombi:2006de} 
one often considers the case in which the local operator is inserted at large distance from the Wilson loop, so that the correlator is used to extract the coefficient of the chiral primary in the operator product expansion of the Wilson loop \cite{Berenstein:1998ij}. However, suppose that we insert the local operator at the north pole of $S^2$. Then we can make a conformal transformation which maps the north pole to infinity, and our results on the matrix model for $O_J$ and $W(C)$ will still hold. Hence we have shown that the 2d YM-conjecture of \cite{Drukker:2007yx,Drukker:2007dw,Drukker:2007qr} on the Wilson loops on $S^2$, refined here to treat the local operators (see also \cite{Pestun:2009nn}), 
implies as a special case the complex matrix models suggested in \cite{Semenoff:2001xp,Semenoff:2006am,Okuyama:2006jc} for the OPE of circular 1/2 (1/4) BPS Wilson loops and chiral primaries.

\section{Strong coupling computations}
\label{string-sec}
The single trace chiral primary operators of ${\cal N}=4$ SYM take the form
\begin{equation}
C_{A_1A_2\cdots A_J} \tr \Phi^{A_1} \Phi^{A_2}\cdots \Phi^{A_J}\,,
\label{CPO}
\end{equation}
where $A_1,\dots,A_J =1,\dots,6$ are $SO(6)_R$ indices, and the tensor $C_{A_1A_2\cdots A_J}$ is symmetric and traceless in each pair of indices. These operators preserve half of the Poincar\'e supersymmetries, and they have protected conformal dimension $\Delta=J$. In particular, an operator of the form
\begin{equation}
\tr \left(u \cdot \Phi \right)^J\,,
\label{CPO2}
\end{equation}
with $u$ a complex six-vector satisfying $u^2=0$, is a chiral primary operator. Note that the operators of interest in this paper, $O_J=\tr \left(\Phi_n+i \Phi_4 \right)^J$, take this form and are therefore chiral primaries. However notice that the complex vector $u$ is taken to be spacetime dependent $u=(x^i/r,i,0,0)$ in this case.

The $AdS_5\times S^5$ dual description of the chiral primary operators is well established from the early days of the $AdS$/CFT correspondence. 
A chiral primary operator of dimension $J$ corresponds in the bulk to the $J$th Kaluza-Klein mode on $S^5$ of a certain supergravity scalar field which is a linear combination of the fluctuations of the metric and RR 4-form potential along the $S^5$ directions\footnote{This description is appropriate as long as $J \ll N$. Chiral primaries of large conformal dimension, $J \sim N$, are dual to spherical $D3$ brane probes wrapping an $S^3$ inside $AdS_5$ or $S^5$. These are the giant gravitons of \cite{McGreevy:2000cw,Hashimoto:2000zp}. For dimension $J \sim N^2$ the most appropriate dual description is given by the backreacted geometries of \cite{Lin:2004nb}.}. This was first worked out in \cite{Lee:1998bxa}, based on the analysis of \cite{Kim:1985ez}, and we refer the reader to the original literature for more details. Let us denote the relevant 10d scalar field by $s(x,y)$, where $x^{\mu}=(z,\vec{x}),\, \vec{x}\in \mathbb{R}^4$ are coordinates on $AdS_5$ with Poincar\'e metric
\begin{equation}
ds^2=\frac{1}{z^2}\left(d\vec{x}^2 +dz^2\right)
\end{equation}
and $y$ are coordinates on $S^5$. Then we can expand the field $s(x,y)$ in $S^5$ spherical harmonics
\begin{equation}
s(x,y)=\sum_J s^J(x)Y^J(y)\,,
\end{equation}
where $Y^J(y)$ are scalar spherical harmonics satisfying
\begin{equation}
\nabla^2_y Y^J = -J(J+4) Y^J\,.
\label{Y-eq}
\end{equation}
Note that spherical harmonics on $S^5$ transform in the same representation of $SO(6)_R$ as the chiral primaries (\ref{CPO}). If we parameterize the $S^5$ in terms of flat $\mathbb{R}^6$ coordinates $\Theta^A$ satisfying $\Theta^2=1$, then the spherical harmonic corresponding to (\ref{CPO}) and satisfying (\ref{Y-eq}) is just
\begin{equation}
Y^J(\Theta)=C_{A_1A_2\cdots A_J} \Theta^{A_1}\Theta^{A_2}\cdots\Theta^{A_J}\,,
\end{equation}
or $Y^J=(u \cdot \Theta)^J$ for the operator in (\ref{CPO2}).
The supergravity equations of motion imply that the 5d scalar fields obey \cite{Lee:1998bxa}\cite{Kim:1985ez} 
\begin{equation}
\nabla^2_x s^J = J(J-4) s^J\,.
\label{sJ-eq}
\end{equation}
By the standard $AdS$/CFT dictionary, a scalar field in $AdS_5$ with $m^2=J(J-4)$ is dual to an operator of conformal dimension $J$, which is identified with the chiral primary operator at the boundary. In the spectrum one finds only modes with $J \ge 2$, consistently with the fact that the dual gauge theory has gauge group $SU(N)$ (chiral primaries with $J=1$ do not exist due to tracelessness of the $SU(N)$ generators).

The equation of motion for the 5d field $s^J(x)$ with a source located at the boundary can be solved in terms of the bulk-to-boundary propagator
\begin{equation}
s^J(z,\vec{x})=\int d^4\vec{x}' G_J(z,\vec{x}; \vec{x}') s^J_0(\vec{x}')\,,
\end{equation}
where $s^J_0$ is the source at the boundary, and the propagator is given by
\begin{equation}
G_J(z,\vec{x}; \vec{x}')={\cal N}_J \frac{z^J}{\left(z^2+(\vec{x}-\vec{x}')^2\right)^J}\,,
\label{propagator}
\end{equation}
where ${\cal N}_J$ is a normalization factor. According to the standard $AdS$/CFT procedure, correlation functions of the chiral primary operators are obtained by plugging the above solution into the supergravity action and functionally differentiating with respect to the source. It is convenient to chose the normalization constant ${\cal N}_J$ so that the chiral primaries corresponding to $s^J_0$ have unit normalized two-point functions. This depends on the absolute normalization of the type IIB supergravity action, see \cite{Lee:1998bxa}, and fixes the normalization constant to be \cite{Berenstein:1998ij}
\begin{equation}
{\cal N}_J=2^{J/2-2} \frac{J+1}{N\sqrt{J}}\,.
\end{equation}
The correspondingly normalized operators on the gauge theory side take the
form\footnote{The perhaps unfamiliar factor of $(-i)^J$ is due to our
  conventions that the fields $\Phi_A$ are anti-Hermitian.} 
\begin{equation}
{\cal O}_J(x)=2^{J/2} (-i)^J \frac{(2\pi)^J}{\lambda^{J/2}\sqrt{J}}\tr \left(u\cdot\Phi\right)^J\,, 
\end{equation}
which satisfy
\begin{equation}
\la {\cal O}_J(x_1) {\cal O}^*_{J'}(x_2)\ra=\frac{\delta_{JJ'}}{\left(x_1-x_2\right)^{2J}}\,.
\end{equation}
Here we have assumed that $u$ is a constant six-vector satisfying $u^2=0$ and $u
\cdot u^*=1$. In the following string calculation we will adopt the same
normalization conventions for our operators $O_J(x)=\tr\left(\frac {x^i}{r} \Phi_i+i\Phi_4\right)^J$ on $S^2$ with $x$-dependent $u$-vector. This means that we consider the normalized operators
\begin{equation}
\tilde{O}_J(x)=2^{J/2} (-i)^J
\frac{(2\pi)^J}{\lambda^{J/2}\sqrt{J}}\tr\left(\frac {x^i}{r} \Phi_i+i\Phi_4\right)^J\,.
\end{equation}
Their two-point function satisfies (note that here we do not want to take the complex conjugate)
\begin{equation}
\begin{aligned}
\la\tilde{O}_J(x_1) \tilde{O}_{J'}(x_2) \ra &=\frac{\delta_{JJ'}}{\left(x_1-x_2\right)^{2J}} 
\left(u(x_1)\cdot u(x_2)\right)^J \\
&=\frac{\delta_{JJ'}}{\left(2(r^2-x_1\cdot x_2)\right)^{J}} 
\left(x_1\cdot x_2/r^2-1\right)^J=\left (\frac {-1}{2 r^2} \right )^J \delta_{JJ'}\,.
\end{aligned}
\end{equation}
From the supergravity point of view, the factor $\left(u(x_1)\cdot u(x_2)\right)^J$ arises from the integration of the two spherical harmonics $(u(x_1) \cdot \Theta)^J$ and $(u(x_2) \cdot \Theta)^{J'}$ over $S^5$, while the denominator comes from the bulk-to-boundary propagator in $AdS_5$.

\subsection{Correlators of local operators and Wilson loops}
\begin{figure}
\begin{center}
\includegraphics[width=85mm]{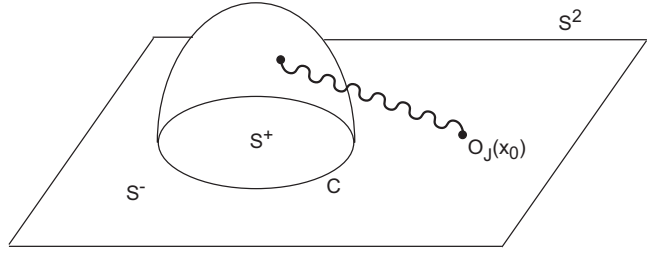}
\parbox{13cm}{\caption{The correlator $\la O_J(x_0)W(C)\ra$ at strong coupling. The plane represents the $S^2\in \mathbb{R}^4$ at the $AdS_5$ boundary, on which the Wilson loop $C$ resides. The operator $O_J(x_0)$ is inserted at a point $x_0\in S^2$, in either of the regions $S^+$ or $S^-$. The wavy line represents the supergravity mode dual to the local operator, which propagates from the insertion point $x_0$ to a point on the string worldsheet dual to the Wilson loop.}
\label{corr-string}}
\end{center}
\end{figure}

In the string theory dual, the correlator between a local operator $O_J(x_0)$ and a Wilson loop is obtained by computing the amplitude for the process in which the supergravity mode dual to $O_J(x_0)$ is emitted from the insertion point $x_0$ at the boundary and then absorbed at a point on the string worldsheet dual to the Wilson loop \cite{Berenstein:1998ij} (see also \cite{Semenoff:2002kk} for a review), see Figure \ref{corr-string}. This requires to work out how the scalar field $s(x,y)$ couples to the string worldsheet.  The Polyakov action for the $AdS_5\times S^5$ string reads, in conformal gauge
\begin{equation}
S=\frac{1}{4\pi\alpha '} \int d^2\sigma \left(g_{\mu\nu}\partial_{\alpha}x^{\mu}\partial_{\alpha}x^{\nu}+g_{mn}\partial_{\alpha}y^m \partial_{\alpha} y^n\right)\,,\quad \mu,\nu=1,\ldots,5\,,\,\,\ m,n=1,\ldots,5\,,
\end{equation} 
where $g_{\mu\nu}$, $g_{mn}$ are the $AdS_5$ and $S^5$ metrics respectively. To linear order, the supergravity mode $s(x,y)$ is related to the metric fluctuations $h_{\mu\nu}$, $h_{mn}$ around the background $g_{\mu\nu}$, $g_{mn}$ as follows \cite{Lee:1998bxa,Berenstein:1998ij,Giombi:2006de}
\begin{equation}
\begin{aligned}
&h_{\mu\nu}=-\frac{6}{5}J s(x,y) g_{\mu\nu}+ \frac{4}{J+1} \nabla_{(\mu}\partial_{\nu)}s(x,y)\,,\\
&h_{mn}=2J s(x,y)g_{mn}\,.
\label{fluct}
\end{aligned}
\end{equation} 
In the first line $\nabla_{(\mu}\partial_{\nu)}$ stands for the symmetric traceless part
\begin{equation}
\nabla_{(\mu}\partial_{\nu)}s(x,y)=\nabla_{\mu}\partial_{\nu}s(x,y)-\frac{1}{5}g_{\mu\nu}\nabla^2_x s(x,y)\,.
\end{equation}
To linear order, the coupling of the $s(x,y)$ field to the worldsheet is simply obtained by inserting the first order fluctuations (\ref{fluct}) into the Polyakov action
\begin{equation}
S^{(1)}=\frac{1}{4\pi\alpha '} \int d^2\sigma \left(h_{\mu\nu}\partial_{\alpha}x^{\mu}\partial_{\alpha}x^{\nu}+h_{mn}\partial_{\alpha}y^m \partial_{\alpha} y^n\right)\,.
\end{equation}
The field $s(x,y)$ can be written in terms of the boundary source as explained in the previous section 
\begin{equation}
s(x,y)=\int d^4\vec{x}' G_J(z,\vec{x};\vec{x}')s^J_0(\vec{x}') Y^J(y)\,,
\label{s-solution}
\end{equation}
and the normalized correlator is then obtained by functionally differentiating with respect to the source
\begin{equation}
\frac{\la \tilde{O}_J(\vec{x}_0) W(C)\ra}{\la W(C)\ra}=-\frac{\delta S^{(1)}}{\delta s^J_0(\vec{x}_0)}\,.
\label{string-correl}
\end{equation} 
The calculation now proceeds analogously to \cite{Berenstein:1998ij,Giombi:2006de,Semenoff:2006am}. One important difference, however, is that in those works the local operator was inserted at large distance from the Wilson loop, which yield a considerable simplification in the form of the bulk-to-boundary propagator. In our case, we wish to insert the local operators on the same $S^2$ on which the Wilson loops live, so we must work exactly in the separation between the local and Wilson operators. 

For the explicit calculation, it will be convenient to work with the flat embedding coordinates $\Theta^A$ on $S^5$ satisfying $\Theta^2=1$. The spherical harmonic corresponding to $O_J(x_0)$ then can be written as 
\begin{equation}
Y^J(\Theta)=\left(u(x_0)\cdot\Theta\right)^J=\left(\frac{x_0^i}{r} \Theta^i+i\Theta^4\right)^J\,, \quad i=1,2,3\,.
\label{YJ}
\end{equation}
The string solutions dual to the 1/8 BPS Wilson loops on $S^2$ reside on a $AdS_3\times S^2$ subspace of $AdS_5\times S^5$, where the $AdS_3\in AdS_5$ is defined by $x^ix^i+z^2=r^2\,,x^4=0$, and the $S^2\in S^5$ is parameterized by the unit 3-vector $\Theta^i$, while $\Theta^{4,5,6}=0$, see \cite{Drukker:2007qr}. 
Hence the $\Theta^4$ in (\ref{YJ}) can be effectively dropped in the following.  

After carrying out the functional differentiation with respect to the source, we get 
\begin{equation}
\begin{aligned}
&\frac{\delta S^{(1)}_{AdS_5}}{\delta s^J_0(\vec{x}_0)}=\frac{1}{4\pi\alpha '}
\int d^2\sigma \left(-2 \frac{J(J-1)}{J+1}G_J(x^{\mu};x_0)\frac{1}{z^2}\partial_{\alpha}x^{\mu}\partial_{\alpha}x^{\mu}
+\frac{4}{J+1}\partial_{\alpha}\partial_{\alpha}G_J(x^{\mu};x_0)\right)Y^J(\Theta)\\
&\frac{\delta S^{(1)}_{S^5}}{\delta s^J_0(\vec{x}_0)}=\frac{1}{4\pi\alpha '} 2J
\int d^2\sigma \,\partial_{\alpha}\Theta^A\partial_{\alpha}\Theta^A \,G_J(x^{\mu};x_0) Y^J(\Theta)\,.
\label{dS1-step1}
\end{aligned}
\end{equation}
To obtain the second term in the first line, we have used the fact that
\begin{equation}
\partial_{\alpha}x^{\mu} \partial_{\alpha} x^{\nu} \nabla_{\mu}\partial_{\nu} G_J=\partial_{\alpha}\partial_{\alpha} G_J\,,
\end{equation}
which can be proven by noting that the string worldsheet satisfies the equations of motion
\begin{equation}
\partial_{\alpha}\partial_{\alpha}x^{\mu}+\Gamma^{\mu}_{\nu\rho}\partial_{\alpha}x^{\nu}
\partial_{\alpha}x^{\rho}=0\,.
\end{equation}
To further simplify (\ref{dS1-step1}), it is convenient to integrate by parts the second term in the first line. This produces a boundary term which can be neglected since $G_J$ goes to zero sufficiently fast at the boundary (at least if the insertion point $x_0$ does not touch the Wilson loop, which we assume to be the case), plus a term proportional to 
\begin{equation}
\partial_{\alpha}\partial_{\alpha} Y^J(\Theta)=\left(J(J-1) \frac{x_0\cdot\partial_{\alpha}\Theta
x_0\cdot\partial_{\alpha}\Theta}{(x_0 \cdot \Theta)^2}-J \partial_{\alpha}\Theta^A\partial_{\alpha}\Theta^A\right) Y^J(\Theta)\,.
\end{equation}
Here we have used the equations of motion for the $\Theta^{A}$ 
\begin{equation}
  \label{eq:theta-a-equation}
  \p_{\alpha} \p_{\alpha} \Theta^{A} + \p_{\alpha} \Theta^{B} \p_{\alpha} \Theta^{B} \Theta^{A} = 0.
\end{equation}

Then the $AdS_5$ and $S^5$ contributions in (\ref{dS1-step1}) can be combined to give our final result
\begin{equation}
\frac{\delta S^{(1)}}{\delta s^J_0(\vec{x}_0)}=\frac{\sqrt{\lambda}}{{2\pi}}\frac{J(J-1)}{J+1}
\int d^2 \sigma \Bigg{(} \partial_{\alpha}\Theta^i\partial_{\alpha}\Theta^j
\left(\delta^{ij}+\frac{2 x_0^i x_0^j}{(x_0\cdot \Theta)^2}\right)-\frac{1}{z^2}\partial_{\alpha}x^{\mu}\partial_{\alpha}x^{\mu}\Bigg{)}G_J Y^J\,,
\label{dS1-final}
\end{equation}
where we have used the relation $\alpha '=1/\sqrt{\lambda}$. By plugging in an explicit string solution and evaluating the above integral one obtains the strong coupling prediction for the correlator (\ref{string-correl}).

\begin{figure}
\begin{center}
\includegraphics[width=70mm]{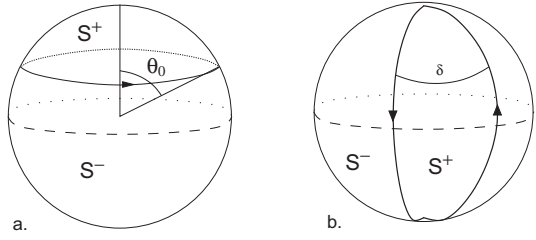}
\parbox{13cm}{\caption{The 1/4 BPS latitude (a) and two-longitudes (b) Wilson loops. The corresponding dual string solutions are given respectively in eq. \ref{latitude} and eq. \ref{longitudes}.}
\label{loops}}
\end{center}
\end{figure}

Let us start by considering the string solution dual to the 1/4 BPS latitude \cite{Drukker:2006ga,Drukker:2007qr}. The corresponding Wilson loop on $S^2$ is depicted in Figure \ref{loops}a. The explicit solution is given by\footnote{Here we consider only the stable solution. One can carry out the analogous calculation for the unstable solution found in \cite{Drukker:2006ga}.}
\begin{equation}
\begin{aligned}
&x^1=r \frac{\tanh\sigma_0 \cos\tau}{\cosh\sigma},\,\quad x^2=r \frac{\tanh\sigma_0 \sin\tau}{\cosh\sigma},\quad x^3=r \frac{1}{\cosh\sigma_0},\quad x^4=0,\\&
z=r \tanh\sigma_0\tanh\sigma\,,\\
&\Theta^1=-\frac{\cos\tau}{\cosh(\sigma_0+\sigma)},\,\qquad \Theta^2=-\frac{\sin\tau}{\cosh(\sigma_0+\sigma)},\,\qquad
\Theta^3=\tanh(\sigma_0+\sigma)\,.
\label{latitude}
\end{aligned}
\end{equation}
Here the range of the coordinates is $0<\sigma<\infty$, $0<\tau<2\pi$, and the parameter $\sigma_0$ is related to the latitude angle on $S^2$ by $\tanh\sigma_0=\sin\theta_0$. Inserting this solution into (\ref{dS1-final}), together with the explicit expression for the propagator given in (\ref{propagator}), one obtains a somewhat complicated expression which we do not explicitly report here. Remarkably, we have verified by direct numerical integration that the result does not depend on the precise position $x_0\in S^2$ of the local operator, but only on whether the operator sits ``inside" or ``outside" the loop. This is precisely the behavior expected from the gauge theory analysis and the relation to 2d YM. The integrand simplifies considerably if we take the insertion point to be the north or south pole: 
$\vec{x_0} = (0,0,\pm r,0)$. By doing a change of variables $\xi=\tanh\sigma$ we obtain 
\begin{equation}
\frac{\delta S^{(1)}}{\delta s^J_0(\vec{x}_0)}=-r^{-J} (\pm 1)^J \frac{2^{-J/2-1}}{N}  
\frac{(J-1) \sqrt{J\lambda} \xi_0^{J+1}}{(1\mp\sqrt{1-\xi_0^2})^J}\int_0^1d\xi \left(\frac{\xi^2+\xi\xi_0}{1+\xi \xi_0}\right)^J \frac{\xi_0(1+\xi^2)+2\xi}{\xi^2(\xi+\xi_0)^2}\,,
\end{equation}
where $\xi_0=\tanh\sigma_0=\sin\theta_0$, and the choice of sign corresponds respectively to the insertion point at the north and south pole. Notice that the $\xi$ integral is convergent since $J\ge 2$, and the integration is elementary
\begin{equation}
\int_0^1d\xi \left(\frac{\xi^2+\xi\xi_0}{1+\xi \xi_0}\right)^J \frac{\xi_0(1+\xi^2)+2\xi}{\xi^2(\xi+\xi_0)^2}=\frac{1}{J-1}\left(\frac{\xi^2+\xi\xi_0}{1+\xi \xi_0}\right)^{J-1}\Bigg{|}_0^1=\frac{1}{J-1}.
\end{equation}
So our final result is
\begin{equation}
\frac{\la \tilde{O}_J(\vec{x}_0) W(C)\ra}{\la W(C)\ra} = \frac{r^{-J}}{N} 2^{-J/2} \sqrt{J \lambda}\begin{cases} 
(\cos\frac{\theta_0}{2})^{J+1} (\sin\frac{\theta_0}{2})^{-J+1}\,,\quad x_0\in S^+\\
(-1)^J (\sin\frac{\theta_0}{2})^{J+1} (\cos\frac{\theta_0}{2})^{-J+1}\,,\quad x_0\in S^- \end{cases}\,.
\end{equation}
We see that this is in precise agreement with the matrix model result (\ref{2MM-strong}), since $A_1/A=\sin^2\frac{\theta_0}{2}$ and $A_2/A=\cos^2\frac{\theta_0}{2}$.

Another string solution which is explicitly known is the one corresponding to the 1/4 BPS loop made out of two half longitudes \cite{Drukker:2007qr}. The corresponding Wilson loop on $S^2$ is depicted in Figure \ref{loops}b. In conformal gauge, the string solution is \cite{Drukker:2007qr}\cite{Giombi:2009ms}
\begin{equation}
\begin{aligned}
&x^1=r\frac{a \sin a\sigma\sin\sigma+\cos a\sigma \cos \sigma}{\cosh\sqrt{1-a^2}\tau}\,, \quad 
x^2=r\frac{a \cos a\sigma\sin\sigma-\sin a\sigma \cos \sigma}{\cosh\sqrt{1-a^2}\tau}\,,\\
&x^3=-r\tanh\sqrt{1-a^2}\tau\,,\quad x^4=0\,,\qquad z=r\frac{\sqrt{1-a^2}\sin\sigma}{\cosh\sqrt{1-a^2}\tau}\,,\\
&\Theta^1=\sin a \sigma\,,\quad \quad \Theta^2 = \cos a\sigma\, \quad \quad
\Theta^3=0\,,
\label{longitudes}
\end{aligned}
\end{equation}
where the range of the coordinates is $-\infty<\tau<\infty$, $0<\sigma<\pi$, and the parameter $a$ is related to the opening angle between the two longitudes by $\delta=\pi (1-a)$. Plugging the explicit solution into (\ref{dS1-final}) we again find that, rather non-trivially, the correlator is (almost) independent of the insertion point $x_0$ as expected, and the final result is
\begin{equation}
\frac{\la \tilde{O}_J(\vec{x}_0) W(C)\ra}{\la W(C)\ra} = \frac{r^{-J}}{N} 2^{-J/2-1} \sqrt{J \lambda}\begin{cases} 
(1+a)^{\frac{J+1}{2}} (1-a)^{\frac{-J+1}{2}}\,,\quad x_0\in S^+\\
(-1)^J (1-a)^{\frac{J+1}{2}} (1+a)^{\frac{-J+1}{2}}\,,\quad x_0\in S^- \end{cases}\,.
\end{equation}
This is again in agreement with the matrix model result, since for this loop one has $A_1/A=(1-a)/2$ and $A_2/A=(1+a)/2$.

\subsection*{Acknowledgments}
The work of S.G. is supported in part by the Center for the
Fundamental Laws of Nature at Harvard University and by NSF grants PHY-024482
and DMS-0244464. The work of V.P. is supported by a Junior Fellowship from the
Harvard Society of Fellows, and grants NSh-3035.2008.2 and RFBR 07-02-00645.

\appendix

\section{Notes on light-cone gauge vs Gaussian matrix models in 2d YM}
\newcommand{\gc}{g_{2d}}
In this appendix we collect some notes about 2d Yang-Mills theory on $S^2$ in the ``Euclidean light-cone gauge'' $A_{\bar z}=0$. In particular, we first re-derive in an alternative way the gauge field propagator by relating it to the two-point function of the field strength. Further, in the next section we directly prove the equivalence between light-cone gauge Feynman diagrams and the two-matrix model for the connected correlator of two latitudes derived in \cite{Giombi:2009ms}\cite{Bassetto:2009rt} from the zero instanton sector of the 2d YM theory.

%Here we collect some relevant notes needed to prove the equivalence of the light-cone gauge and hermitian %multi-matrix model
%for the zero instanton sector of the 2d YM. 

All formulas below are only about the 2d theory, so in this appendix 
we do not have to distinguish the 2d fields from the 4d fields and we do not write tilde for the 2d fields. The 2d coupling constant is denoted by $\gc$. For simplicity, we will also take the $S^2$ to have unit radius. 

We use complex coordinates $z, \bar z$ on $S^2$. The metric for radius $r=1$ takes the form
\begin{equation}
  \label{eq:metric-S^2}
  ds^2 = \frac {4 dz d \bar z} { (1 + z \bar z)^2},
\end{equation}
so we have 
\begin{equation}
  \label{eq:metric-g_z-zbar}
  g_{z \bar z} = g_{\bar z z} = \frac {2} { (1 + z \bar z)^2},\quad
 g_{zz} = 0,\quad g_{\bar z \bar z} = 0.
\end{equation}
The volume form on $S^2$ is 
\begin{equation}
  \label{eq:volume-form}
\mu  = -i d\bar z \wedge dz g_{\bar z z} = \frac { 4 d^2 x }{(1 + |z|^2)^2}\,,
\end{equation}
where $z = x_1 + i x_2$ and $d^2 x = dx_1 \wedge dx_2 = -\frac i 2 d \bar z \wedge dz = -\frac i 2 d^2 z$.

The 2d YM action in the gauge $A_{\bar z} = 0$  (we skip the Lie algebra indices assuming contractions
where needed)  
\begin{equation}
  \label{eq:2dym-action}
  S =  \frac {1} {2 \gc^2} \int \mu F_{\bar z z } F^{\bar z z}
\end{equation}
explicitly takes the form 
\begin{equation}
  \label{eq:S-explicit}
  S = - \frac {1} {2 \gc^2} \int -i d \bar z \wedge d z F_{\bar z z}^2 (2\rho^2)^{-1} = 
-\frac {1 } {2 \gc^2} \int d^2 x \rho^{-2} F_{\bar z z}^2 \,,
\end{equation}
where we have introduced
\begin{equation}
  \label{eq:rho-definition}
  \rho = \frac {1} { 1 + z \bar z}.
\end{equation}

We now wish to represent the correlation functions of $A_{z}$ in terms 
of correlation functions of $F_{\bar z z} = \pbar A_{z}$. 

We can change variables in the path integral from $A_{z}$ to $\pbar A_{z}$. The Jacobian of this change of variables 
is trivial, but in the integration domain over $F_{\bar z z}$ we need to explicitly project out
the zero modes $F_{\bar z z} d \bar z \wedge d z = c \mu$ where $c$ is a constant, because 
such modes are not in the image of $\bar \p: \Omega^{1,0} \to \Omega^{1,1}$.
Hence, from the free action (\ref{eq:S-explicit}) we immediately get the correlation function of
the free fields $F_{\bar z z}$
\begin{equation}
  \label{eq:correlation-F}
 \langle F_{\bar z z}(x) F_{\bar z z}(x') \rangle = -\gc^2 \left(  \rho(x)^2 \delta^2 (x-x') -\frac 1 {\int d^2 x \rho(x)^2 } \rho(x)^2 \rho(x')^2 \right),
\end{equation}
which, of course, agrees\footnote{In our conventions
$\int d^2 z \delta^2(z) = 1$, and $ \int d^2 x \delta^2(x) = 1$, and $d^2 z = d \bar z \wedge d z = 2 i d^2 x$ hence $\delta^2(z) = \frac 1 {2 i} \delta^2 (x)$ for $z=x_1 + i x_2$.}
 with (\ref{prop-2d}) since
\begin{equation}
  \int d^2 x \rho(x)^2 = {\pi}.
\end{equation}

Next we express $A_{z}$ in terms of $F_{z \bar z}$. Using the fact that
\begin{equation}
 \p_{\bar z} \frac {1} {z} = \pi \delta^2 (x) 
\end{equation}
one easily gets
\begin{equation}
\label{eq:A-in-F}
  A_{z}(z) = \frac {1} {2 \pi i} \int  d^2 u   \frac 
{F_{\bar u u}(u)} {z - u}.
\end{equation}
The relation (\ref{eq:A-in-F}) makes sense on $\CP^{1}$ if $\int d^2 u F_{\bar u u} = 0$.

Now, using (\ref{eq:A-in-F}) and (\ref{eq:correlation-F}) we represent the propagator for $A_{z}$ by means
of auxiliary integrals over $u$-planes
\begin{equation}
  \label{eq:A-z-propagator-integrals}
  \la A_{z} (z) A_{w} (w)\ra  = \frac {1} {(2 \pi i)^2} \int d^2 u \int d^2 u' \frac {1} {z - u} \frac {1} {w - u'} 
\la F_{\bar u u}(u)  F_{\bar u u} (u') \ra 
\end{equation}

The $\delta$-function term in (\ref{eq:correlation-F}) removes one integral and for the remaining integration we use
\begin{equation}
  \label{eq:identity}
  \int \frac{d^2 u} {2i} \frac {1} { (z-u)(w-u)} \frac {1} {(1+|u|^2)^2}  = - \pi \frac {1}{ (1+ |z|^2)(1+|w|^2)} \frac { \bar z - \bar w}
{z - w} + \pi \frac { \bar w} {(1+|w|^2)} \frac {\bar z} {(1+|z|^2)}.
\end{equation}
This identity can be shown by doing the integral over circles $|u|=\const$ using residues and then integrating 
over $|u|$.
The contribution of the second term in (\ref{eq:correlation-F}) is obtained using the integral 
\begin{equation}
  \label{eq:one-integral-rho}
 \frac 1 {{ 2 \pi i}} \int  d^2 u \frac{ \rho(u) } {z-u}  = \frac {\bar z} { 1 + |z|^2}.
\end{equation}
Hence, the contribution of the second term in (\ref{eq:correlation-F}) is precisely cancelled by 
the second term in (\ref{eq:identity}) and we get
\begin{equation}
  \label{eq:A-propagator}
  \la A_z (z) A_{w} (w) \ra = \frac {\gc^2}{\pi} \frac {1}{ 1+ |z|^2} \frac {1} {1 + |w|^2} \frac { \bar z - \bar w}
{ z - w},
\end{equation}
which, of course, agrees with (\ref{light-prop}).

\subsection{Connected correlator of two circular Wilson loops on $S^2$}
\newcommand{\tz}{\tilde z}
Here we explicitly compute the Feynman diagrams for the connected correlator of two latitude Wilson loops on $S^2$ using the propagator (\ref{eq:A-propagator}) for the gauge fields. We prove directly the equivalence of the light-cone gauge with the Hermitian two-matrix model of \cite{Giombi:2009ms}\cite{Bassetto:2009rt}.
 
We consider two concentric contours, the first contour $C_1$ given by
$|z| = r_1$ and the second contour $C_2$ given by $|z| = r_2$. We assume that $r_1 < r_2$. 

We denote points on $C_1$ as  $w_{i}$ and points on $C_2$ as $z_{i}$. There are three types of propagators (here we use the relations $\bar w_{i} = r_1^2 /w_i$ and $\bar z_i = r_2^2 /z_i$):

From $C_1$ to $C_1$ contour:
\begin{equation}
  \label{eq:c1-c1}
  \la A_{w}(w_i) dw_{i} A_{w}(w_j) dw_{j} \ra = \frac{ \gc^2}{\pi} \frac {- r_1^2 }{(1 + r_1^2)^2} \frac{ dw_i}{w_i} \frac{dw_j}{w_j}\,.
\end{equation}

From $C_2$ to $C_2$ contour:
\begin{equation}
  \label{eq:c2-c2}
  \la A_{z}(z_i) dz_{i} A_{z}(z_j) dz_{j} \ra = \frac{ \gc^2}{\pi} \frac {- r_2^2 }{(1 + r_2^2)^2} \frac{ dz_i}{z_i} \frac{dz_j}{z_j}\,.
\end{equation}

From $C_2$ to $C_1$ contour:
\begin{equation}
  \label{eq:C2-to-C1}
\begin{split}
  \la A_{z}(z) dz A_{w}(w) dw \ra = \frac { \gc^2}{\pi} \frac {-r_1^2} { (1+r_1^2)} \frac {1}{(1+r_2^2)}
\left ( \frac {z- r_2^2/r_1^2 w  } { wz (z - w)}  \right) dz \, dw   = \\
= \frac {-r_1^2} { (1+r_1^2)} \frac {1}{(1+r_2^2)} \left (  \frac {1}{wz} - \frac {r_2^2/r_1^2 -1}{wz} \sum_{n=1}^\infty 
\lb \frac w z \rb^n \right) dz \, dw\,.
\end{split}
\end{equation}

\begin{figure}
\begin{center}
\includegraphics[width=50mm]{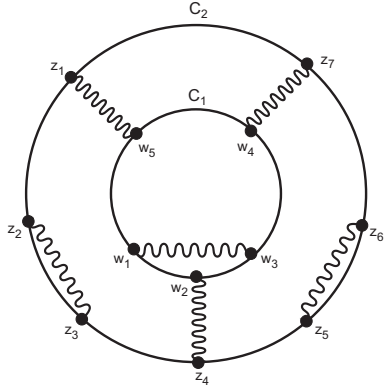}
\parbox{13cm}{\caption{A typical ladder diagram for the connected correlator of two circular Wilson loops on $S^2$. In this example, the points $z_1,z_4,z_7$ belong to $T(2,1)$ and $w_5,w_2,w_4$ are the corresponding points in $I(T(2,1))$. The points $z_2,z_3,z_5,z_6$ belong to $T(2,2)\cup I(T(2,2))$, and finally $w_1,w_3$ are in $T(1,1)\cup J(T(1,1))$.}
\label{latitude-fig}}
\end{center}
\end{figure}

Now consider a ladder Feynman diagram where we take $n$ points on the contour $C_2$ and
$n'$ points on the contour $C_1$. A typical diagram is depicted in Figure \ref{latitude-fig}. There are two type of  points $z_k$. The points $z_k$ of the first type
are connected by propagators to points $w_{I(k)}$ on the contour $C_1$, 
where $I(k)$ labels the point which connects with $z_k$. 
The points of the second type on $C_2$ are pairwise connected with each other.
We denote the set of the first type as $T(2,1)$ and the set of the second type as a disjoint union 
of $T(2,2)$ and $I(T(2,2))$. In other words, the set $T(2,2)$ contains a half of the points of the second type, and the set $I(T(2,2))$ contains the remaining half. 
Let $T_{21}$ be the number of points in $T(2,1)$, 
and $T_{22}$ be the number of points in $T(2,2)$. 
Analogously, the $T(1,1)$ denotes the points
on $C_1$ connected to the points in $J(T(1,1))$ on $C_1$, and the connection map is denoted by $J(k)$, 
i.e. we say that a point $w_{k}$ for $k \in T(1,1)$ 
connects to a point $w_{J(k)}$. Clearly, $n + n' = 2( T_{22} + T_{21} + T_{11}) $.

As usual, these Feynman diagrams arise from expanding each Wilson loop in the correlator $\la W_R(C_2)W_{R'}(C_1) \ra $ in powers of the gauge field 
\begin{equation}
  \label{eq:expansion}
  W_{R}(C) = 1 + \sum_{n=1}^{\infty} W_{R}^{(n)}(C),
\end{equation}
and taking all Wick contractions with the gauge field propagator. For example the term with $n$ points on $C_2$ is 
\begin{equation}
  \label{eq:powers}
 W_{R}^{(n)}(C_2) =  \dimR^{-1} \int_{r_2}^{e^{2 \pi i}r_2} dz_{n} \cdots
\int_{r_2}^{z_{k+1}} dz_{k} \cdots \int_{r_2}^{z_{2}} dz_{1} \tr_{R} A_{z}(z_1) \cdots A_{z}(z_n),
\end{equation}
where here and in all formulae below symbol $\int_{a}^{b} dz$ means integration over the contour $C_2$ such
that the points $a,z,b$ on $C_2$ are placed in the counterclockwise order.
Using the cyclic invariance of the trace we can rewrite this term as the $1/n$-th of all cyclic permutations of the points $z_1, \dots, z_n$ on the contour, and we get
\begin{equation}
  \label{eq:cyclic-invariance}
\begin{split}
 W_{R}^{(n)}(C_2) = \dimR^{-1} \frac {1}{n} \int_{r_2}^{e^{2\pi i}r_2} dz_{1} \int_{z_1}^{e^{2 \pi i}z_1} dz_{n} \cdots
\int_{z_1}^{z_{k+1}} dz_{k} \cdots \int_{z_1}^{z_{3}} dz_{2} \tr_{R} \left ( A_{z}(z_1) \cdots A_{z}(z_n) \right ) = \\
= \frac {1}{n} \int_{r_2}^{e^{2\pi i}r_2} dz_{1} \int_{1}^{e^{2 \pi i}} z_1 d \tz_{n} \cdots
\int_{1}^{\tz_{k+1}} z_1 d \tz_{k} \cdots \int_{1}^{\tz_{3}} z_1 d \tz_{2} \tr_{R} \left (A_{z}(z_1) 
A_{z}(z_1 \tz_2) \cdots A_{z}(z_1 \tz_n) \right),
\end{split}
\end{equation}
where in the last line we have changed integration variables $z_k = z_1 \tz_{k}$ for $k=2\dots n$.
  
Now, using (\ref{eq:cyclic-invariance}) and (\ref{eq:c1-c1})(\ref{eq:c2-c2})(\ref{eq:C2-to-C1}), we can explicitly evaluate any given Feynman diagram $D$, which defines for us the sets $T(2,2), T(2,1), T(1,1), I(k), J(k)$ introduced above.  For each point $z_k \in C_2$ for $k>1$ we substitute our change of variables $z_{k} = z_1 \tilde z_{k}$. Without loss of generality, let us suppose that the point $z_1$ belongs to the set $T(2,1)$\footnote{Since we are computing the connected correlator, $T(2,1)$ is non-empty and we can always use cyclic invariance so that $z_1\in T(2,1)$. In any case, it will be clear from the computation that it does not matter to which set $z_1$ belongs.}. Then for the given ladder diagram $D$ we get  
\begin{multline}
\label{eq:expl-corr-integral}
\la W^{(n)}_R(C_2) W^{(n')}_{R'}(C_1) \ra_D =\\=  \frac{1}{\dimR \dimRpr}\lb \frac {\gc^2}{\pi} \rb^{(n+n')/2} \lb \frac {-r_2^2}{(1+r_2^2)^2} \rb^{T_{22}}  \lb \frac {-r_1^2} { 1+r_1^2} \frac {1}{ 1+ r_2^2} \rb ^{T_{21}} 
\lb \frac { -r_1^2 } {(1 + r_1^2)^2} \rb^{T_{11}}
\times \\
\int_{r_1}^{e^{2 \pi i}r_1} d w_{n'} \int_{r_1}^{w_{n_1}-1} d w_{n' -1}\int_{r_1}^{w_2} d w_{1}  \frac {1}{n} \int_{r_2 }^{e^{2\pi i} r_2} dz_{1} \int_{1}^{e^{2 \pi i}} z_1 d \tz_{n} \cdots
\int_{1}^{\tz_{k+1}} z_1 d \tz_{k} \cdots \int_{1}^{\tz_{3}} z_1 d \tz_{2}  \times \\
\times 
\left ( \frac {z_1- r_2^2/r_1^2 w_{I(1)}  } { w_{I(1)} z_{1} (z_1 - w_{I(1)} )}  \right)
\left ( \prod_{\substack{k \in T(2,1) \\ k \neq 1}}  \frac {z_1 \tz_k- r_2^2/r_1^2 w_{I(k)}  } { w_{I(k)} z_1 \tz_{k} (z_1 \tz_k - w_{I(k)} )}  \right)
\left ( \prod_{k \in T(2,2)}  \frac {1 } {  z_1 \tz_{k} z_1 \tz_{I(k)} }  \right)
\left ( \prod_{k \in T(1,1)}  \frac {1 } {  w_{k} w_{J(k)} }  \right) \times \\
\times 
\delta_{i_1 j_{I(1)}} \prod_{\substack{ k \in T(2,1)\\ k \neq 1}} \delta_{i_k j_{I(k)}} 
\prod_{k \in T(2,2)} \delta_{i_k i_{I(k)}}
\prod_{k \in T(1,1)} \delta_{j_k j_{J(k)}}
\tr_{R} T_{i_1} \cdots T_{i_n} \tr_{R'} T_{j_1} \cdots T_{j_{n'}}\,.
\end{multline}

Now is the key step of the computation. First we evaluate the contour integral over $z_1$. The
integrand in (\ref{eq:expl-corr-integral}) is a rational function with respect to $z_1$, so we can evaluate the integral by residues.
Using that $|w_i| < |z_1|$ and that $|\tz_k| = 1$, we can see that there are no residues outside the integration contour $|z_1| =r_2$, except the residue at $z_1 = \infty$. So, taking the residue at $z_1 = \infty$ we get
\begin{multline}
\label{eq:expl-corr-integral-after-z1}
\la W^{(n)}_R(C_2) W^{(n')}_{R'}(C_1) \ra_D =\\= \frac{1}{\dimR \dimRpr} (2\pi i) \lb \frac {\gc^2}{\pi} \rb^{(n+n')/2} \lb \frac {-r_2^2}{(1+r_2^2)^2} \rb^{T_{22}}  \lb \frac {-r_1^2} { 1+r_1^2} \frac {1}{ 1+ r_2^2} \rb ^{T_{21}} 
\lb \frac { -r_1^2 } {(1 + r_1^2)^2} \rb^{T_{11}}
\times \\
\int_{r_1}^{e^{2 \pi i}r_1} d w_{n'} \int_{r_1}^{w_{n_1}-1} d w_{n' -1}\int_{r_1}^{w_2} d w_{1}  
\frac {1}{n} 
% \int_{r_2 }^{e^{2\pi i r_2}} dz_{1} %
 \int_{1}^{e^{2 \pi i}}  d \tz_{n} \cdots
\int_{1}^{\tz_{k+1}}  d \tz_{k} \cdots \int_{1}^{\tz_{3}}  d \tz_{2}  \times \\
\times \frac {1} {\tz_2 \tz_3 \dots \tz_{n}} 
\times \frac {1} {w_{1} w_{2} \dots w_{n'}}\\
%\left ( \frac {z_1- r_2^2/r_1^2 w_{I(1)}  } { w_{I(1)} z_{1} (z_1 - w_{I(1)} )}  \right)
%\left ( \prod_{\substack{k \in T(2,1) \\ k \neq 1}}  \frac {z_1 \tz_k- r_2^2/r_1^2 w_{I(k)}  } { w_{I(k)} z_1 \tz_{k} (z_1 \tz_k - w_{I(k)} )}  \right)
%\left ( \prod_{k \in T(2,2)}  \frac {1 } {  z_1 \tz_{k} z_1 \tz_{I(k)} }  \right)
%\left ( \prod_{k \in T(1,1)}  \frac {1 } {  w_{k} w_{J(k)} }  \right) \times \\
\times 
\delta_{i_1 j_{I(1)}} \prod_{\substack{ k \in T(2,1)\\ k \neq 1}} \delta_{i_k j_{I(k)}} 
\prod_{k \in T(2,2)} \delta_{i_k i_{I(k)}}
\prod_{k \in T(1,1)} \delta_{j_k j_{J(k)}}
\tr_{R} T_{i_1} \cdots T_{i_n} \tr_{R'} T_{j_1} \cdots T_{j_{n'}}\,.
\end{multline}
The remaining integrations are now elementary. The integral over $\tz_2,\dots, \tz_n$ gives a factor $\frac {(2 \pi i)^{n-1}} {(n-1)!}$ 
and the integral over $w_1,\dots,w_{n'}$ gives a factor $\frac {(2\pi i)^{n'}} {n'!}$. Hence, we finally get
\begin{multline}
\label{eq:expl-corr-integral-after-z1-result}
\la W^{(n)}_R(C_2) W^{(n')}_{R'}(C_1) \ra_D =\\= \frac{1}{\dimR \dimRpr} (2\pi i)^{n+n'} \frac {1}{n! n'!} \lb \frac {\gc^2}{\pi} \rb^{(n+n')/2} \lb \frac {-r_2^2}{(1+r_2^2)^2} \rb^{T_{22}}  \lb \frac {-r_1^2} { 1+r_1^2} \frac {1}{ 1+ r_2^2} \rb ^{T_{21}} 
\lb \frac { -r_1^2 } {(1 + r_1^2)^2} \rb^{T_{11}}
\times \\
\delta_{i_1 j_{I(1)}} \prod_{\substack{ k \in T(2,1)\\ k \neq 1}} \delta_{i_k j_{I(k)}} 
\prod_{k \in T(2,2)} \delta_{i_k i_{I(k)}}
\prod_{k \in T(1,1)} \delta_{j_k j_{J(k)}}
\tr_{R} T_{i_1} \cdots T_{i_n} \tr_{R'} T_{j_1} \cdots T_{j_{n'}}\,.
\end{multline}

This expression agrees with the corresponding Feynman diagram in the two-matrix model \cite{Giombi:2009ms}\cite{Bassetto:2009rt}. To see
the equivalence one needs expressions for the areas $A_1,A_2$ on $S^2$ written in terms of $r_1, r_2$. We denote by $A_1$ the area of the disk on $S^2$ inside $C_1$ ($|z| < r_1$) and by $A_2$ the area outside $C_2$ ($|z| > r_2$). Using
$z = e^{i\phi} \tan \frac \theta 2$ we have
\begin{equation}
  \label{eq:trigon}
  \frac {r_{1}^2}{ 1 + r_1^2} = \frac{ A_{1}}{A}, \quad
  \frac {1}{1 + r_1^2} = \frac  { A - A_{1} }{A}, \quad
  \frac {r_{2}^2}{ 1 + r_2^2} =  \frac { A - A_{2}}{A}, \quad
  \frac {1}{1 + r_2^2} = \frac {A_{2}}{A}\,.
\end{equation}

Then 
\begin{multline}
\label{eq:expl-corr-integral-after-z1-result-siml}
\la W^{(n)}_R(C_2) W^{(n')}_{R'}(C_1) \ra_D =\\= \frac{1}{\dimR \dimRpr}  \frac {1}{n! n'!} \lb  4 \pi \gc^2 \rb^{(n+n')/2} \lb 
\frac {(A-A_2)A_2}{A^2} \rb^{T_{22}}  \lb \frac {A_1 A_2} { A^2} \rb ^{T_{21}} 
\lb \frac { (A-A_1)A_1 } {A^2} \rb^{T_{11}}
\times \\
 \prod_{\substack{ k \in T(2,1)}} \delta_{i_k j_{I(k)}} 
\prod_{k \in T(2,2)} \delta_{i_k i_{I(k)}}
\prod_{k \in T(1,1)} \delta_{j_k j_{J(k)}}
\tr_{R} T_{i_1} \cdots T_{i_n} \tr_{R'} T_{j_1} \cdots T_{j_{n'}}\,.
\end{multline}

This precisely agrees with the corresponding Feynman diagram in the two-matrix model \cite{Giombi:2009ms}\cite{Bassetto:2009rt}
\begin{equation}
  \label{eq:two-matrix-model}
  \int DX DY \exp \lb  -\frac 1 {2 \gc^2} \lb \frac {X_a X_a}{A_1} + \frac { (X_a - Y_a)(X_a - Y_a)}{A-A_1-A_2} 
+ \frac { Y_a Y_a }{ A_{2}} \rb \rb \frac {1} { \dimRpr} \tr_{R'} e^{T_a X_a} \frac{1}{\dimR} \tr_{R} e^{T_a Y_a},
\end{equation}
which has been derived from the zero-instanton sector of 2d YM. 

\bibliography{bsample}
\end{document}